\documentclass[a4paper]{article}

\usepackage{a4wide}
\usepackage{graphicx}
\usepackage{psfrag}
\usepackage{citesort}
\usepackage{amsmath}
\usepackage{amssymb}

\renewcommand\Re{\mathop{\rm Re}}
\renewcommand\Im{\mathop{\rm Im}}
\newcommand\Res{\mathop{\rm Res}\nolimits}
\newcommand\Eta{{\mathrm{H}}}
\newcommand\aL{a_{\mathrm{L}}}
\newcommand\aR{a_{\mathrm{R}}}
\newcommand\bL{b_{\mathrm{L}}}
\newcommand\bR{b_{\mathrm{R}}}
\newcommand\dL{d_{\mathrm{L}}}
\newcommand\dR{d_{\mathrm{R}}}
\newcommand\fL{f_{\mathrm{L}}}
\newcommand\fR{f_{\mathrm{R}}}
\newcommand\gL{g_{\mathrm{L}}}
\newcommand\gR{g_{\mathrm{R}}}
\newcommand\nL{n_{\mathrm{L}}}
\newcommand\nR{n_{\mathrm{R}}}
\newcommand\tL{t_{\mathrm{L}}}
\newcommand\tR{t_{\mathrm{R}}}
\newcommand\FL{F_{\mathrm{L}}}
\newcommand\FR{F_{\mathrm{R}}}
\newcommand\RL{R_{\mathrm{L}}}
\newcommand\RR{R_{\mathrm{R}}}
\newcommand\WL{W_{\mathrm{L}}}
\newcommand\WR{W_{\mathrm{R}}}
\newcommand\betaL{\beta_{\mathrm{L}}}
\newcommand\betaR{\beta_{\mathrm{R}}}
\newcommand\muL{\mu_{\mathrm{L}}}
\newcommand\muR{\mu_{\mathrm{R}}}
\newcommand\rhoL{\rho_{\mathrm{L}}}
\newcommand\rhoR{\rho_{\mathrm{R}}}
\newcommand\phiL{\phi_{\mathrm{L}}}
\newcommand\phiR{\phi_{\mathrm{R}}}
\newcommand\SigmaL{\Sigma_{\mathrm{L}}}
\newcommand\SigmaR{\Sigma_{\mathrm{R}}}
\newcommand\rhotd{\rho_{\triangledown}}

\title{Bethe Ansatz solution of triangular trimers on the triangular
  lattice}
\author{Alain Verberkmoes and Bernard Nienhuis\\
  \sl Instituut voor Theoretische Fysica, Valckenierstraat~65,\\
  \sl 1018~XE~Amsterdam, The~Netherlands}
\date{September 4, 1999}

\begin{document}

\maketitle

\begin{abstract}
\noindent
Details are presented of a recently announced exact solution of a model
consisting of triangular trimers covering the triangular lattice.  The
solution involves a coordinate Bethe Ansatz with two kinds of
particles.  It is similar to that of the square--triangle random tiling
model, due to M.~Widom and P.~A.~Kalugin.  The connection of the trimer
model with related solvable models is discussed.
\end{abstract}

\section{Introduction}

The dimer problem is one of the classic models of statistical
mechanics.  A dimer in this context is a particle that occupies two
neighbouring sites of a lattice.  In the dimer--monomer model dimers
and monomers (particles occupying one lattice site each) are placed on
a lattice such that they cover all sites, without overlap.
Equivalently the monomers can be viewed as empty sites; the lattice is
then partly covered with dimers.
This model was introduced to describe diatomic molecules
adsorbed on a substrate~\cite{fowler:1937}.  Attempts have been made in
vain to solve this model exactly, that is, to calculate its free
energy.  The special case that there are no empty sites (monomers) is
known as the dimer problem.  There the dimers cover the lattice
completely and without overlap.  This model has been solved for planar
lattices, independently by Kasteleyn~\cite{kasteleyn:1961} and by
Temperley and Fisher~\cite{temperley:1961}.  Their solution is based on
the possibility to express the partition function of the model as a
Pfaffian.  For many planar lattices the dimer problem can also be
solved by means of the Bethe Ansatz.  On the honeycomb lattice for
example
it can be formulated as a five-vertex model.  This is a special case of
the six-vertex model, whose Bethe Ansatz solution is
well-known~\cite{lieb:1967,lieb:1967a,lieb:1967b,lieb:1967c,%
sutherland:1967,yang:1967,sutherland:1967a,lieb:1972}.  A review of the
dimer problem is given in~\cite{nagle:1989}.

\begin{figure}[!hbt]
  \hfil\includegraphics[scale=0.4]{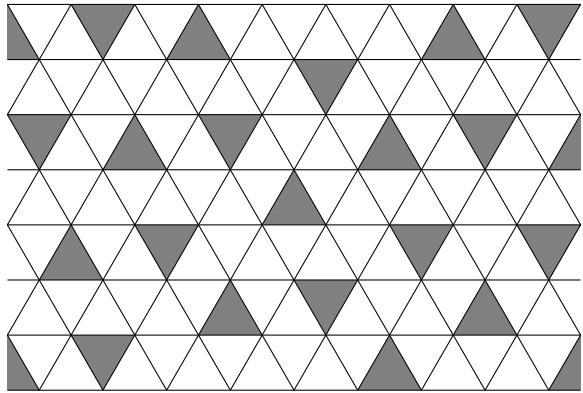}\hfil
  \caption{A typical configuration of the trimer model.  Each lattice
    site belongs to precisely one trimer.}
  \label{fig:trimers:config}
\end{figure}

Inspired by the solvability of the dimer model, we consider lattice
coverings by trimers.  A trimer is a particle that occupies three
lattice sites.  We only consider triangular trimers, which live
naturally on the triangular lattice.  As in the dimer model, we require
that these particles cover the lattice completely and without overlap.
Thus every lattice site is covered by precisely one trimer.
Figure~\ref{fig:trimers:config} shows a typical configuration.

As will be shown in Subsection~\ref{sub:trimers:domainwalls} this model
has a structure of domains separated by domain walls.  The domains are
hexagonal, and the domain walls form a honeycomb network.  Similar
domain wall structures are used to describe an incommensurate phase of
a monolayer of a monoatomic gas adsorbed on a hexagonal
substrate~\cite{villain:1980}.  The entropy of such a network is
largely due to the ``breathing'' of the cells: it is possible to blow
up a domain and simultaneously shrink its six neighbours, or vice
versa.

Hexagonal domain wall structures also occur in the square--triangle
random tiling model~\cite{kawamura:1983}.  For that model a coordinate
Bethe Ansatz was found by Widom~\cite{widom:1993}.  The resulting Bethe
Ansatz equations were solved analytically in the thermodynamic limit by
Kalugin~\cite{kalugin:1994}.  An exact solution of the trimer model was
announced in~\cite{verberkmoes:1999b}; in the present paper we describe
its the derivation.  The solution is very similar to that for the
square--triangle tiling, and we closely follow Kalugin's arguments.
The outline of our calculation is as follows.  A transfer matrix for
the model is formulated.  After the choice of a reference state two
types of elementary excitations are found.  They are closely related to
the above-mentioned domain wall structure of the model.  In order to
diagonalise the transfer matrix a coordinate Bethe Ansatz is set up in
terms of the elementary excitations.  The resulting semi-grand
canonical ensemble is discussed.  In the thermodynamic limit the Bethe
Ansatz leads to a set of two coupled integral equations.  These can be
solved in a special case.  From their solution the relevant physical
quantities are computed.  The results of the calculation are summarised
in Subsection~\ref{sub:trimers:summary}.  We then consider the entropy
as function of the density of down trimers.  The model undergoes
two phase transitions in the density of down trimers.

Finally we discuss the relation between the trimer model, the
square--triangle random tiling model, and yet another solvable model
with a hexagonal domain wall structure.

\section{Preliminaries}

\subsection{Sub-lattices}
\label{sub:trimers:sublat}

\begin{figure}[!hbt]
  \psfrag{0}[B][B][1][0]{0}
  \psfrag{1}[B][B][1][0]{1}
  \psfrag{2}[B][B][1][0]{2}
  \psfrag{3}[B][B][1][0]{3}
  \psfrag{4}[B][B][1][0]{4}
  \psfrag{5}[B][B][1][0]{5}
  \hfil\includegraphics[scale=0.4]{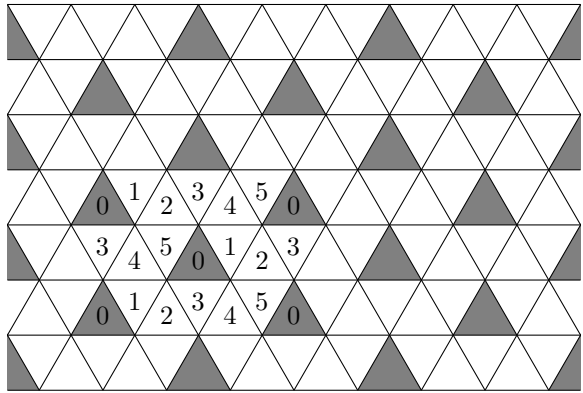}\hfil
  \caption{A regular configuration in which the trimers occupy a
    sub-lattice of the faces.  There are six such sub-lattices,
    numbered 0, 1, \dots,~5.}
  \label{fig:trimers:reference}
\end{figure}

Figure~\ref{fig:trimers:reference} shows a very regular configuration
of the model, in which the trimers are positioned on a sub-lattice of
the triangular faces.  There are six such sub-lattices, which we number
0, 1, \dots, 5 as indicated in the figure.  Note that the even-numbered
sub-lattices consist of the up triangles while the down triangles
constitute the odd-numbered sub-lattices.  For a given configuration
let $N$ denote the total number of trimers and let $N_i$ denote the
number of trimers on sub-lattice~$i$.  We wish to compute the entropy
per trimer as a function of the sub-lattice densities
\begin{displaymath}
  \rho_0 = \frac{N_0}{N}, \qquad
  \rho_1 = \frac{N_1}{N}, \qquad
  \dots, \qquad
  \rho_5 = \frac{N_5}{N}.
\end{displaymath}
These densities satisfy the obvious linear constraint
\begin{equation}
  \rho_0 + \rho_1 + \rho_2 +
  \rho_3 + \rho_4 + \rho_5 = 1.
  \label{equ:trimers:linear}
\end{equation}
In Subsection~\ref{sub:trimers:constraint} it will be shown that when
toroidal boundary conditions are imposed the densities also satisfy a
quadratic constraint:
\begin{equation}
  \rho_0 \rho_2 + \rho_2 \rho_4 + \rho_4 \rho_0 =
  \rho_1 \rho_5 + \rho_3 \rho_5 + \rho_5 \rho_1.
  \label{equ:trimers:quadratic}
\end{equation}
Hence of the six sub-lattice densities only four are independent.  In
order to be able to set up a transfer matrix we pass to the grand
canonical ensemble.  The trimers on each sub-lattice~$i$ are given a
fugacity $w_i = \exp(\mu_i)$.  After the transfer matrix has been
diagonalised we shall Legendre transform back to the canonical
ensemble.

\subsection{Domains and walls}
\label{sub:trimers:domainwalls}

\begin{figure}[!hbt]
  \psfrag{1}[B][B][1][0]{1}
  \psfrag{2}[B][B][1][0]{2}
  \psfrag{3}[B][B][1][0]{3}
  \psfrag{5}[B][B][1][0]{5}
  \hfil\includegraphics[scale=0.4]{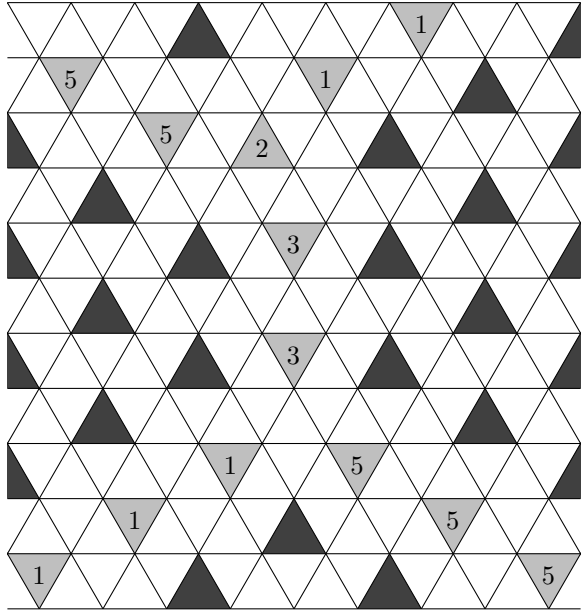}\hfil
  \caption{The configuration from
    Figure~\protect\ref{fig:trimers:reference} admits line excitations.
    These domain walls can meet in Ys (top) and upside-down Ys
    (bottom).  The Ys are chiral; the mirror image of the Y shown here
    contains a trimer on sub-lattice 4 instead of~2.  The
    upside-down Ys are achiral.  To guide the eye the trimers
    not on sub-lattice~0 are shaded lighter; the numbers indicate their
    sub-lattices.}
  \label{fig:trimers:collision}
\end{figure}

Occupying sub-lattice~0 completely while leaving the other five
sub-lattices empty results in the configuration of the model shown in
Figure~\ref{fig:trimers:reference}.  This arrangement does not admit
local changes.  However, it is possible to flip a whole line of
trimers.  Such a line can be viewed as a wall separating two domains
consisting of trimers on sub-lattice~0.  These domain walls come in
three types (orientations), corresponding to the three odd-numbered
sub-lattices.  When two walls of different type meet a wall of the
third type is formed.  A trimer on sub-lattice 2 or 4 occurs when three
domain walls of different type meet in a Y, but this does not happen at
an upside-down~Y.  Figure~\ref{fig:trimers:collision} show examples of
how the three types of domain walls can meet.  In a general
configuration the domain walls form a hexagonal network.

\subsection{Transfer matrix}

In an allowed configuration of the model each lattice site belongs to
precisely one trimer.  This trimer sits either on one of the three
lattice faces above the site or on one of the three faces below the
site.  Label the site with a ``spin'' $\uparrow$ or $\downarrow$
accordingly.

Consider a horizontal row of lattice sites and assume that the trimer
configuration below that row is given.  It determines the spins on that
row.  The sites occupied by a trimer below have a spin $\downarrow$,
while those not occupied by such a trimer must needs carry a
spin~$\uparrow$.  Now consider the next layer of lattice faces, above
this row.  In order to decide what trimer configurations on this layer
are possible, it is sufficient to know which sites are already covered.
This is precisely the information encoded by the spins.

This shows that the model can be described
in terms of a transfer matrix that connects two consecutive rows of
spins.  Let $\boldsymbol\sigma$ denote be the spin configuration on the
lower row and $\boldsymbol\tau$ the spin configurations on the upper
row.  Consider all trimer arrangements (without overlaps) on the layer
in between that are compatible with the spin configurations
$\boldsymbol\sigma$ and~$\boldsymbol\tau$.  (Generally there is at most
one such arrangement.)  The sum of their Boltzmann weights is the
transfer matrix element~$T_{\boldsymbol{\tau\sigma}}$.

The rows of lattice sites in the model come in two types, call them A
and B, that are shifted with respect to each other.  The rows of type A
and B alternate.  Hence there are in fact two transfer matrices,
$T_{\mathrm{AB}}$ for layers with upper row of type A and lower row
of type B, and $T_{\mathrm{BA}}$ for layers with upper row of type B
and lower row of type~A\@.  The products
$T_{\mathrm{AB}} T_{\mathrm{BA}}$ and
$T_{\mathrm{BA}} T_{\mathrm{AB}}$ are double transfer matrices
that act between rows of equal type.
Take a lattice consisting of $2M$ layers and impose periodic boundary
conditions in the vertical direction, that is, identify the lower and
upper row.  As usual the partition function of the model on this
lattice is
\begin{displaymath}
  Z = \mathop{\rm Tr} \, (T_{\mathrm{AB}} T_{\mathrm{BA}})^M =
      \mathop{\rm Tr} \, (T_{\mathrm{BA}} T_{\mathrm{AB}})^M.
\end{displaymath}
In the limit that $M$ tends to infinity the partition sum is dominated
by
the largest eigenvalue of $T_{\mathrm{AB}} T_{\mathrm{BA}}$ or
$T_{\mathrm{BA}} T_{\mathrm{AB}}$.

The matrices $T_{\mathrm{AB}}$ and $T_{\mathrm{BA}}$ can be
combined into a single matrix
\begin{displaymath}
  T = \begin{pmatrix}
        0 & T_{\mathrm{AB}} \\
        T_{\mathrm{BA}} & 0
      \end{pmatrix}
\end{displaymath}
acting on vectors
\begin{displaymath}
  \psi =
  \begin{pmatrix}
    \psi_{\mathrm{A}} \\ \psi_{\mathrm{B}}
  \end{pmatrix}
\end{displaymath}
where $\psi_{\mathrm{A}}$ and $\psi_{\mathrm{B}}$ are ``wave
functions'' on the rows of type A and B, respectively.  If such a
vector is an eigenvector of $T$ with eigenvalue $\Lambda$ then
$\psi_{\mathrm{A}}$ and $\psi_{\mathrm{B}}$ are eigenvectors
of $T_{\mathrm{AB}} T_{\mathrm{BA}}$ and
$T_{\mathrm{BA}} T_{\mathrm{AB}}$, respectively, with
eigenvalue~$\Lambda^2$.

Some identification of the row types A and~B could have been chosen in
order to avoid the complication discussed.  This amounts to skewing the
lattice or, equivalently, the transfer matrix direction.  Because it
breaks a mirror symmetry of the system we have avoided this solution.

\subsection{Conserved quantities and elementary excitations}

\begin{figure}[!hbt]
  \psfrag{0}[B][B][1][0]{0}
  \psfrag{1}[B][B][1][0]{1}
  \psfrag{2}[B][B][1][0]{2}
  \psfrag{3}[B][B][1][0]{3}
  \psfrag{4}[B][B][1][0]{4}
  \psfrag{5}[B][B][1][0]{5}
  \hfil\includegraphics[scale=0.4]{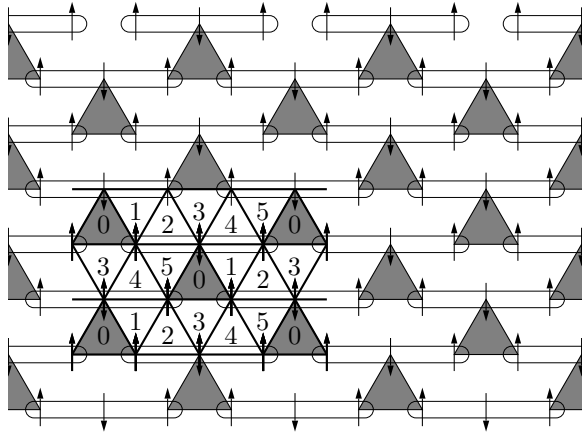}\hfil
  \caption{Spins for the configuration from
    Figure~\protect\ref{fig:trimers:reference}.  For clarity the edges
    of the triangular lattice have been largely omitted.}
  \label{fig:trimers:blocks}
\end{figure}

In the configuration obtained by fully occupying sub-lattice~0, each
row of spins consists of repeating blocks $\uparrow\downarrow\uparrow$.
Therefore we group the sites into blocks of three, as in
Figure~\ref{fig:trimers:blocks}.  Number the blocks in a row from the
left to the right.

Consider a trimer configuration on a layer of the lattice.  Let $L$
denote the number of blocks per row and let $n_0$, $n_1$, \dots, $n_5$
denote the number of trimers on each sub-lattice.  The horizontal
and vertical lattice direction are viewed as ``space'' and ``time'',
respectively; the lower and upper row of the layer then are
time-slices at times $t$ and~$t+1$.
Counting the number
of $\uparrow$ spins in the lower row, distinguishing by the position
inside the block, gives:
\begin{align*}
  n_{\uparrow \bullet \bullet}^{(t)} &= n_0 + n_1 + n_2, \\
  n_{\bullet \uparrow \bullet}^{(t)} &= n_2 + n_3 + n_4, \\
  n_{\bullet \bullet \uparrow}^{(t)} &= n_4 + n_5 + n_0.
\end{align*}
From this we get
\begin{align}
  n_{\downarrow \bullet \bullet}^{(t)} +
  n_{\bullet \uparrow \bullet }^{(t)} &=
  L - n_0 - n_1 + n_3 + n_4,
  \label{equ:trimers:nllower} \\
  n_{\bullet \bullet \downarrow}^{(t)} +
  n_{\bullet \uparrow \bullet}^{(t)} &=
  L - n_0 + n_2 + n_3 - n_5.
  \label{equ:trimers:nrlower}
\end{align}
Counting the number of $\downarrow$ spins in the upper row gives:
\begin{align*}
  n_{\downarrow \bullet \bullet}^{(t+1)} &= n_3 + n_4 + n_5, \\
  n_{\bullet \downarrow \bullet}^{(t+1)} &= n_5 + n_0 + n_1, \\
  n_{\bullet \bullet \downarrow}^{(t+1)} &= n_1 + n_2 + n_3.
\end{align*}
From this we get
\begin{align}
  n_{\downarrow \bullet \bullet}^{(t+1)} +
  n_{\bullet \uparrow \bullet}^{(t+1)} &=
  L - n_0 - n_1 + n_3 + n_4,
  \label{equ:trimers:nlupper} \\
  n_{\bullet \bullet \downarrow}^{(t+1)} +
  n_{\bullet \uparrow \bullet}^{(t+1)} &=
  L - n_0 + n_2 + n_3 - n_5.
  \label{equ:trimers:nrupper}
\end{align}

Comparing (\ref{equ:trimers:nllower}) with (\ref{equ:trimers:nlupper})
and (\ref{equ:trimers:nrlower}) with (\ref{equ:trimers:nrupper}) we
see that the quantities
$n_{\downarrow \bullet \bullet} + n_{\bullet \uparrow \bullet}$ and
$n_{\bullet \bullet \downarrow} + n_{\bullet \uparrow \bullet}$
are conserved between rows.

These conserved quantities are nonnegative.  The only row of spins for
which both are zero consists entirely of blocks
$\uparrow\downarrow\uparrow$.  There is only one way to fit a layer of
trimers above this row.  Of course the row of spins above that layer
consists
again entirely of blocks $\uparrow\downarrow\uparrow$.  This
row state will be chosen as the ``empty'' or reference state for the
Bethe Ansatz in Section~\ref{sec:trimers:betheansatz}.

\begin{figure}[!hbt]
  \psfrag{5}[B][B][1][0]{5}
  \hfil\includegraphics[scale=0.4]{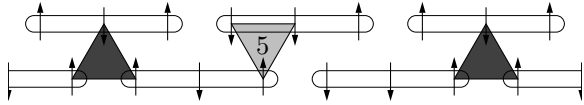}\hfil
  \caption{There is only one way to fit trimers below a row consisting
    of one block $\downarrow\downarrow\uparrow$ amidst blocks
    $\uparrow\downarrow\uparrow$.  It leads to another such row of
    spins below.}
  \label{fig:trimers:ddutransfer}
\end{figure}

A row of spins with
$n_{\downarrow \bullet \bullet} + n_{\bullet \uparrow \bullet} = 1$ and
$n_{\bullet \bullet \downarrow} + n_{\bullet \uparrow \bullet} = 0$ is
obtained by replacing one block, say at position $x$, in the reference
state with $\downarrow\downarrow\uparrow$.
There is only one possible
configuration of trimers on the layer above, see
Figure~\ref{fig:trimers:ddutransfer}.  The row above consists of blocks
$\uparrow\downarrow\uparrow$ except for one block
$\downarrow\downarrow\uparrow$ at position $x-\frac{1}{2}$.
Thus the transfer matrix has
shifted the block $\downarrow\downarrow\uparrow$ in the lower row half
a step to the left in the upper row.  This block is a left-moving
elementary excitation of the reference state.  It will be called an
L-particle.
Similarly the block $\uparrow\downarrow\downarrow$ is an elementary
right-moving excitation, or R-particle.
The conserved quantities
$n_{\downarrow \bullet \bullet} + n_{\bullet \uparrow \bullet}$ and
$n_{\bullet \bullet \downarrow} + n_{\bullet \uparrow \bullet}$
are the number $\nL$ of L-particles and the
number $\nR$ of R-particles, respectively.

The particle content of the blocks $\uparrow\downarrow\uparrow$,
$\downarrow\downarrow\uparrow$ and $\uparrow\downarrow\uparrow$ has now
been determined.  For each of the other five blocks both $\nL$ and
$\nR$ are greater than zero.  Therefore these blocks are combinations
of the elementary excitations.  They will be discussed in more detail
in Subsections~\ref{sub:trimers:oneone}--\ref{sub:trimers:anyany}.

We have found no other conserved quantities than $\nL$ and $\nR$
(except in the special case when $\nL = 0$ or $\nR = 0$).

\subsection{World lines and quadratic constraint}
\label{sub:trimers:constraint}

\begin{figure}[!hbt]
  \hfil\includegraphics[scale=0.4]{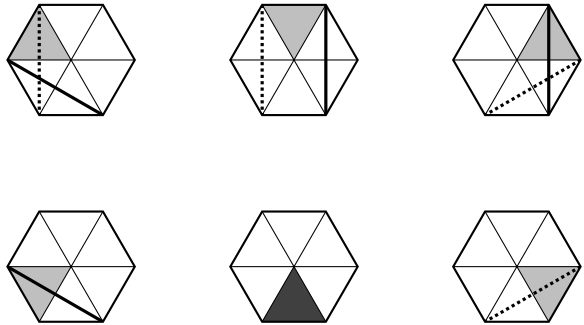}\hfil
  \caption{Decompose the triangular lattice into hexagonal patches such
    that the lower middle triangle of each patch belongs to
    sub-lattice~0.  The other triangles, in counterclockwise order,
    then belong to sub-lattice 1, 4, 3, 2 and 5.  These
    patches can be decorated with the world lines
    of the L-particles (solid) and R-particles (dashed).}
  \label{fig:trimers:patches}
\end{figure}

Divide the lattice into hexagonal patches containing one face from each
sub-lattice, in such a way that the lower middle triangle of each patch
belongs to sub-lattice~0.  There are six
trimer configurations possible on such a patch.  Decorate each patch
according to this configuration as shown in
Figure~\ref{fig:trimers:patches}.  It is tedious but straightforward
to verify
that the decorations of the patches making up the lattice fit together,
giving a set of solid and dashed lines running from the bottom to the
top of the lattice.
It can also be checked that the crossings of these lines with the
lattice rows correspond to the locations of the L-particles (solid
lines) and R-particles (dashed lines).
Hence these lines are the ``world lines'' of the L-particles
and R-particles where the horizontal and vertical lattice direction are
viewed as ``space'' and ``time'' respectively.

Impose toroidal boundary conditions.
We now derive the quadratic constraint~(\ref{equ:trimers:quadratic}) by
the same method as that used for rectangle--triangle random tiling
models in \cite{gier:1997a} and~\cite{gier:1998}.
Cut the torus open along a
horizontal row of sites, so that the model is now on a cylinder.  By
stacking a number of copies of the configuration on top of each other
we can achieve that each world line winds around the cylinder an
integer number of times.  Let $2M$ be the number of rows.
In each row we can count the number of L-particles and R-particles:
\begin{align*}
  \nL &= L - n_0 - n_1 + n_3 + n_4, \\
  \nR &= L - n_0 + n_2 + n_3 - n_5.
\end{align*}
Summing over the entire lattice yields
\begin{align}
  2 M \nL &= 2 L M - N_0 - N_1 + N_3 + N_4,
  \label{equ:trimers:totalnleft} \\
  2 M \nR &= 2 L M - N_0 + N_2 + N_3 - N_5.
  \label{equ:trimers:totalnright}
\end{align}
Since the world lines of the L-particles do not cross each other, they
all have
the same winding number~$\WL$.  In order to compute this
winding number consider the total leftward movement of the L-particles.
There are $\nL$ such particles, each of them winds around
the cylinder $\WL$ times, and each winding constitutes a
movement of $L$ blocks to the left, so the total leftward movement
amounts $\nL \WL L$ blocks.  It can be seen
from Figure~\ref{fig:trimers:patches} that an L-particle moves half
a block to the left at a trimer on sub-lattice 2 or~5, while there is
no horizontal movement of L-particles at a trimer on another
sub-lattice.  Summing over the entire lattice shows the total leftward
movement of the L-particles to be $\frac{1}{2}(N_2+N_5)$ blocks.
Hence
\begin{equation}
  \nL \WL L =
  \frac{1}{2} \left( N_2 + N_5 \right).
  \label{equ:trimers:windleft}
\end{equation}
Fully analogously one has
\begin{equation}
  \nR \WR L =
  \frac{1}{2} \left( N_1 + N_4 \right).
  \label{equ:trimers:windright}
\end{equation}
Now consider the number of crossings of L-particle lines and R-particle
lines.  There are $\nL$ L-particles each winding leftward
around the cylinder $\WL$ times, and $\nR$
R-particles each winding rightward around the cylinder
$\WR$ times, so the number of crossings is
$\nL \nR
(\WL+\WR)$.
From Figure~\ref{fig:trimers:patches} it is seen that crossings
occur precisely at trimers on sub-lattice 2 or~4, so that the number of
crossings is $N_2+N_4$.  Equating these two expressions for the
number of crossings gives
\begin{displaymath}
  \nL \nR
  (\WL+\WR) = N_2+N_4.
\end{displaymath}
Substituting into the above equation first
(\ref{equ:trimers:windleft}) and
(\ref{equ:trimers:windright}), then (\ref{equ:trimers:totalnleft}) and
(\ref{equ:trimers:totalnright}), then multiplying by $2LM$ and finally
using
\begin{displaymath}
  2 L M = N_0 + N_1 + N_2 + N_3 + N_4 + N_5
\end{displaymath}
yields
\begin{displaymath}
  N_1 N_3 + N_3 N_5 + N_5 N_1 =
  N_0 N_2 + N_2 N_4 + N_4 N_0.
\end{displaymath}
Dividing by $N^2$ gives~(\ref{equ:trimers:quadratic}).  For the sake of
the argument we have stacked a number of copies of the original
configuration on top of each other, but the resulting
(\ref{equ:trimers:quadratic})
is not affected by this.

\section{Bethe Ansatz}
\label{sec:trimers:betheansatz}

In this section we describe a Bethe Ansatz (BA) that diagonalises the
transfer matrix.  Since the particle numbers $\nL$ and
$\nR$ are conserved quantities the transfer matrix is
block diagonal.  We begin by considering the sector with
$\nL=0$ and $\nR=0$ and then pass to sectors
with higher particle numbers.

\subsection{No particles}
\label{sub:trimers:zerozero}

The only state in the sector $\nL=0$, $\nR=0$
is the reference state which consists entirely of blocks
$\uparrow\downarrow\uparrow$, so this sector is one-dimensional.
Therefore the transfer matrix acting on this sector is trivially
diagonal.
The layer between two consecutive rows in the reference state consists
of $L$ trimers on sub-lattice~0, so its Boltzmann weight is $w_0^L$.
It is the eigenvalue of the transfer matrix in this sector.
For convenience we define the ``reduced'' transfer matrix $\tilde T$ to
be the transfer matrix $T$ divided by~$w_0^L$.

\subsection{One L-particle}

Consider a row of spins containing a single L-particle
($\downarrow\downarrow\uparrow$) at position~$x$.  The transfer matrix
has shifted this particle from position $x+\frac{1}{2}$ in the row
below half a step to the left, see
Figure~\ref{fig:trimers:ddutransfer}.  The layer between the two rows
contains $L-1$ trimers on sub-lattice~0 and one trimer on
sub-lattice~5.  Hence the action of the (reduced) transfer matrix on
the ``wave function'' is given by
\begin{displaymath}
  (\tilde T \psi)(
  \downarrow \downarrow \uparrow x) =
  \frac{w_5}{w_0} \psi(
  \downarrow \downarrow \uparrow x + {\textstyle \frac{1}{2}}).
\end{displaymath}
(We use $\downarrow\downarrow\uparrow x$ as notation for the
row configuration that has a block $\downarrow\downarrow\uparrow$ at
position $x$ and blocks $\uparrow\downarrow\uparrow$ at the other
positions.)
The solution of the eigenvalue problem
$\tilde T \psi = \tilde\Lambda\psi$ is
\begin{displaymath}
  \psi(
  \downarrow \downarrow \uparrow x) =
  A_u u^x,
\end{displaymath}
where $A_u$ is some constant, and
\begin{displaymath}
  \tilde\Lambda = \frac{w_5}{w_0} u^{\frac{1}{2}}.
\end{displaymath}

\subsection{One L-particle and one R-particle}
\label{sub:trimers:oneone}

Consider a row of spins containing an L-particle
($\downarrow\downarrow\uparrow$) at position~$x$ and an R-particle
($\uparrow\downarrow\downarrow$) at position~$y$, with~$x<y$.  If the
particles are apart, this situation was formed by shifting the
L-particle to the left and the R-particle to the right:
\begin{equation}
  (\tilde T \psi)(
  \downarrow \downarrow \uparrow x,
  \uparrow \downarrow \downarrow y) =
  \frac{w_5 w_1}{w_0^2} \psi(
  \downarrow \downarrow \uparrow x + {\textstyle \frac{1}{2}},
  \uparrow \downarrow \downarrow y - {\textstyle \frac{1}{2}})
  \quad \mbox{if $y-x > 1$}.
  \label{equ:trimers:oneone1}
\end{equation}
(We write the arguments of $\psi$ in order of increasing position; for
example, the notation in the LHS of (\ref{equ:trimers:oneone1}) implies
that~$x<y$.)
If however the particles are next to each other, the situation was
formed from a ``bound state'' ($\downarrow\downarrow\downarrow$), see
Figure~\ref{fig:trimers:collision} (top):
\begin{equation}
  (\tilde T \psi)(
  \downarrow \downarrow \uparrow z - {\textstyle \frac{1}{2}},
  \uparrow \downarrow \downarrow z + {\textstyle \frac{1}{2}}) =
  \frac{w_5 w_1}{w_0^2} \psi(
  \downarrow \downarrow \downarrow z).
  \label{equ:trimers:oneone2}
\end{equation}
This bound state was formed from another type of bound state
($\uparrow\uparrow\uparrow$):
\begin{equation}
  (\tilde T \psi)(
  \downarrow \downarrow \downarrow z) =
  \frac{w_4 w_1}{w_0} \psi(
  \uparrow \uparrow \uparrow z - {\textstyle \frac{1}{2}}) +
  \frac{w_5 w_2}{w_0} \psi(
  \uparrow \uparrow \uparrow z + {\textstyle \frac{1}{2}})
  \label{equ:trimers:oneone3}.
\end{equation}
The two terms correspond to two chiral configurations one of which is
depicted in Figure~\ref{fig:trimers:collision}.  This bound state was
formed from an R-particle and an L-particle next to each other, the
R-particle sitting to the left of the L-particle:
\begin{equation}
  (\tilde T \psi)(
  \uparrow \uparrow \uparrow z) =
  \frac{1}{w_0} \psi(
  \uparrow \downarrow \downarrow z - {\textstyle \frac{1}{2}},
  \downarrow \downarrow \uparrow z + {\textstyle \frac{1}{2}}).
  \label{equ:trimers:oneone4}
\end{equation}
This configuration can have arisen from the same bound state again.
The alternation of this bound state and the situation where the
R-particle and L-particle are next to each other
($\uparrow\downarrow\downarrow\ \downarrow\downarrow\uparrow$)
corresponds to the vertical domain wall in
Figure~\ref{fig:trimers:collision}.
The configuration where the particles are next to each other can also
have arisen by shifting the R-particle half a step to the right and the
L-particle half a step to the left, see
Figure~\ref{fig:trimers:collision} (bottom):
\begin{equation}
  (\tilde T \psi)(
  \uparrow \downarrow \downarrow z - {\textstyle \frac{1}{2}},
  \downarrow \downarrow \uparrow z + {\textstyle \frac{1}{2}}) =
  w_3 \psi(\uparrow \uparrow \uparrow z) +
  \frac{w_1 w_5}{w_0^2} \psi(
  \uparrow \downarrow \downarrow z - 1,
  \downarrow \downarrow \uparrow z + 1).
  \label{equ:trimers:oneone5}
\end{equation}
Finally, a configuration where the particles are apart was formed by
shifting the R-particle half a step to the right and the L-particle
half a step to the left:
\begin{equation}
  (\tilde T \psi)(
  \uparrow \downarrow \downarrow y,
  \downarrow \downarrow \uparrow x) =
  \frac{w_1 w_5}{w_0^2} \psi(
  \uparrow \downarrow \downarrow y - {\textstyle \frac{1}{2}},
  \downarrow \downarrow \uparrow x + {\textstyle \frac{1}{2}})
  \quad \mbox{if $x-y > 1$}.
  \label{equ:trimers:oneone6}
\end{equation}
We want to solve the eigenvalue equation
$\tilde T \psi = \tilde\Lambda \psi$ for
(\ref{equ:trimers:oneone1})--(\ref{equ:trimers:oneone6}).
The eigenvalue equation for (\ref{equ:trimers:oneone1}) and
(\ref{equ:trimers:oneone2}) is satisfied by
\begin{align*}
  \psi(
  \downarrow \downarrow \uparrow x,
  \uparrow \downarrow \downarrow y) &=
  A_{u v} u^x v^y, \\
  \psi(
  \downarrow \downarrow \downarrow z) &=
  A_{u v} u^z v^z
\end{align*}
with eigenvalue
\begin{displaymath}
  \tilde\Lambda =
  \frac{w_5}{w_0} u^{\frac{1}{2}} \;
  \frac{w_1}{w_0} v^{-\frac{1}{2}}.
\end{displaymath}
Similarly the eigenvalue equation for (\ref{equ:trimers:oneone4}),
(\ref{equ:trimers:oneone5}), and (\ref{equ:trimers:oneone6}), with the
same value for~$\tilde\Lambda$, is satisfied by
\begin{align*}
  \psi(
  \uparrow \downarrow \downarrow y,
  \downarrow \downarrow \uparrow x) &=
  \phantom{D B_{v u}} A_{v u} v^y u^x
  \quad \mbox{if $x-y > 1$}, \\
  \psi(
  \uparrow \downarrow \downarrow z - {\textstyle \frac{1}{2}},
  \downarrow \downarrow \uparrow z + {\textstyle \frac{1}{2}}) &=
  \phantom{D} B_{v u} A_{v u} v^{z-\frac{1}{2}} u^{z+\frac{1}{2}},
  \\
  \psi(
  \uparrow \uparrow \uparrow z) &=
  D B_{v u} A_{v u} v^z u^z,
\end{align*}
where
\begin{align}
  B_{v u} &=
  \left(1 - \frac{w_0^3 w_3}{w_1^2 w_5^2} u^{-1} v\right)^{-1},
  \label{equ:trimers:b} \\
  D &= \frac{w_0}{w_1 w_5}.
  \label{equ:trimers:d}
\end{align}
The eigenvalue equation for (\ref{equ:trimers:oneone3}) is satisfied
too if
\begin{displaymath}
  \frac{A_{u v}}{A_{v u}} = S_{u v},
\end{displaymath}
where
\begin{equation}
  S_{u v} =
  \frac{w_0^2}{w_1 w_5} \,
  \left(\frac{w_4}{w_5} u^{-1} + \frac{w_2}{w_1} v\right)
  \left(1 - \frac{w_0^3 w_3}{w_1^2 w_5^2} u^{-1} v\right)^{-1}.
  \label{equ:trimers:suv}
\end{equation}
The above analysis suggests to interpret the bound state
$\downarrow\downarrow\downarrow$ as LR (in that order) and the bound
state $\uparrow\uparrow\uparrow$ as~RL\@.  The eigenfunction is then
written
\begin{align*}
  \psi(\mathrm{L} \ x, \mathrm{R} \ y) &= A_{u v} u^x v^y,
  \\[5 pt]
  \psi(\mathrm{R} \ y, \mathrm{L} \ x) &=
  \begin{cases}
    \phantom{D B_{v u}} A_{v u} v^y u^x & \mbox{if $x-y \ge 2$},
    \\
    \noalign{\vskip 3 pt}
    \phantom{D} B_{v u} A_{v u} v^y u^x & \mbox{if $x-y = 1$}, \\
    \noalign{\vskip 3 pt}
    D B_{v u} A_{v u} v^y u^x & \mbox{if $x-y = 0$}.
  \end{cases}
\end{align*}

\subsection{Two L-particles and one R-particle}

A similar but more tedious analysis can be carried out for the sector
with two L-particles and one R-particle.  There is a new bound state
($\downarrow\uparrow\uparrow$) that can be interpreted as~LRL\@.  A
solution of the eigenvalue problem
$\tilde T \psi = \tilde\Lambda \psi$ is given by
\begin{align*}
  \psi(\mathrm{L} \ x_1, \mathrm{L} \ x_2, \mathrm{R} \ y) &=
  \textstyle
  \sum_{\pi} A_{u_{\pi(1)} u_{\pi(2)} v}
             u_{\pi(1)}^{x_1} u_{\pi(2)}^{x_2} v^y,
  \\[5 pt]
  \psi(\mathrm{L} \ x_1, \mathrm{R} \ y, \mathrm{L} \ x_2) &=
  \begin{cases}
    \sum_{\pi} \phantom{D B_{v u_{\pi(2)}}}
               A_{u_{\pi(1)} v u_{\pi(2)}}
               u_{\pi(1)}^{x_1} v^y u_{\pi(2)}^{x_2} &
    \mbox{if $x_2-y \ge  2$}, \\
    \noalign{\vskip 3 pt}
    \sum_{\pi} \phantom{D} B_{v u_{\pi(2)}}
               A_{u_{\pi(1)} v u_{\pi(2)}}
               u_{\pi(1)}^{x_1} v^y u_{\pi(2)}^{x_2} &
    \mbox{if $x_2-y = 1$}, \\
    \noalign{\vskip 3 pt}
    \sum_{\pi} D B_{v u_{\pi(2)}} A_{u_{\pi(1)} v u_{\pi(2)}}
               u_{\pi(1)}^{x_1} v^y u_{\pi(2)}^{x_2} &
    \mbox{if $x_2-y = 0$},
  \end{cases}
  \\[5 pt]
  \psi(\mathrm{R} \ y, \mathrm{L} \ x_1, \mathrm{L} \ x_2) &=
  \begin{cases}
    \sum_{\pi} \phantom{D B_{v u_{\pi(1)}}}
               A_{v u_{\pi(1)} u_{\pi(2)}}
               v^y u_{\pi(1)}^{x_1} u_{\pi(2)}^{x_2} &
    \mbox{if $x_1-y \ge 2$}, \\
    \noalign{\vskip 3 pt}
    \sum_{\pi} \phantom{D} B_{v u_{\pi(1)}}
               A_{v u_{\pi(1)} u_{\pi(2)}}
               v^y u_{\pi(1)}^{x_1} u_{\pi(2)}^{x_2} &
    \mbox{if $x_1-y = 1$}, \\
    \noalign{\vskip 3 pt}
    \sum_{\pi} D B_{v u_{\pi(1)}} A_{v u_{\pi(1)} u_{\pi(2)}}
               v^y u_{\pi(1)}^{x_1} u_{\pi(2)}^{x_2} &
    \mbox{if $x_1-y = 0$},
  \end{cases}
\end{align*}
where $\pi$ runs through the permutations of $\{1, 2$\}.  The
amplitudes must satisfy
\begin{alignat}{3}
  \frac{A_{u_i u_{i'} v}}{A_{u_{i'} u_i v}} &=
  \frac{A_{v u_i u_{i'}}}{A_{v u_{i'} u_i}} &&= -1 &
  \quad (i \ne {i'}),
  \label{equ:trimers:asuu} \\
  \frac{A_{u_i v u_{i'}}}{A_{v u_i u_{i'}}} &=
  \frac{A_{u_{i'} u_i v}}{A_{u_{i'} v u_i}} &&= S_{u_i v} &
  \quad (i \ne {i'}),
  \label{equ:trimers:asuv}
\end{alignat}
with $S_{u_i v}$ given by~(\ref{equ:trimers:suv}).  Note that the
amplitude ratio in (\ref{equ:trimers:asuu}) does not depend on $v$ and
that the amplitude ratio in (\ref{equ:trimers:asuv}) does not depend
on~$u_{i'}$.  The eigenvalue is given by
\begin{displaymath}
  \tilde\Lambda =
  \frac{w_5}{w_0} u_1^{\frac{1}{2}} \;
  \frac{w_5}{w_0} u_2^{\frac{1}{2}} \;
  \frac{w_1}{w_0} v^{-\frac{1}{2}}.
\end{displaymath}

\subsection{Arbitrary particle numbers}
\label{sub:trimers:anyany}

With two L-particles and two R-particles, there is a new bound state
($\downarrow\uparrow\downarrow$) that can be interpreted as~LRLR\@.
This completes the list of possible blocks and their interpretation in
terms of particles, see Table~\ref{tab:trimers:blocks}.

\begin{table}[!hbt]
  \caption{The three-spin blocks.}
  \label{tab:trimers:blocks}
  \hfil
  \begin{tabular}{||c||c||}
  \hline
  spins & particles \\
  \hline
  $\uparrow\downarrow\uparrow$     & none \\
  $\downarrow\downarrow\uparrow$   & L \\
  $\uparrow\downarrow\downarrow$   & R \\
  $\downarrow\downarrow\downarrow$ & LR \\
  $\uparrow\uparrow\uparrow$       & RL \\
  $\downarrow\uparrow\uparrow$     & LRL \\
  $\uparrow\uparrow\downarrow$     & RLR \\
  $\downarrow\uparrow\downarrow$   & LRLR \\
  \hline
  \end{tabular}
  \hfil
\end{table}

The solution given above of the eigenvalue problem
$\tilde T \psi = \tilde\Lambda \psi$ for two L-particles and one
R-particle generalises to the higher sectors.
Before describing this generalisation we introduce a notational
convention.  The index $i$, running from 1 to $\nL$, will be
used to number L-particle positions and Bethe Ansatz.  The index $j$,
between 1 and $\nR$, will refer to R-particles.
Now consider a succession of
L-particles with coordinates
$x_1 \le x_2 \le \dots \le x_{\nL}$ and
R-particles with coordinates
$y_1 \le y_2 \le \dots \le y_{\nR}$.  (Note that
$x_i = x_{i+1}$ can arise only from a block LRL or LRLR, so
$x_i = y_j = x_{i+1}$ for some~$y_j$.)  The value of the
wave function is given by
\begin{equation}
  \psi(\mbox{particle sequence}) =
  \sum_{\pi} \sum_{\sigma}
  \prod \left( D \mbox{ and } B_{\dots} \right) A_{\dots}
  \prod \left( u_{\pi(i)}^{x_i} \mbox{ and } v_{\sigma(j)}^{y_j}
        \right),
  \label{equ:trimers:psigeneral}
\end{equation}
where $\pi$ and $\sigma$ run through all permutations of
$\{1, 2, \dots, \nL\}$ and
$\{1, 2, \dots, \nR\}$, respectively.  We shall now
describe the factors in the RHS of~(\ref{equ:trimers:psigeneral}).
For each segment $\mathrm{R} \ y_j, \mathrm{L} \ x_i$ in the
particle sequence with $x_i-y_j = 1$ there is a
factor~$B_{v_{\sigma(j)} u_{\pi(i)}}$.
For each such segment with $x_i-y_j = 0$ there is a
factor~$D B_{v_{\sigma(j)} u_{\pi(i)}}$.
The amplitude $A_{\dots}$ depends on the sequence of the variables
$u$ and $v$
corresponding to the sequence of L-particles and R-particles.  The $u$
are in the order
$u_{\pi(1)}, u_{\pi(2)}, \dots, u_{\pi(\nL)}$ and
the $v$ are in the order $v_{\sigma(1)}, v_{\sigma(2)}, \dots,
v_{\sigma(\nR)}$, but the two sequences interlace.  The
amplitudes $A_{\dots}$ are defined up to an overall factor by the
conditions
\begin{alignat*}{2}
  \frac{A_{\dots u_i u_{i'} \dots}}{A_{\dots u_{i'} u_i \dots}}
  &= -1 && \quad (i \ne {i'}), \\
  \frac{A_{\dots v_j v_{j'} \dots}}{A_{\dots v_{j'} v_j \dots}}
  &= -1 && \quad (j \ne {j'}), \\
  \frac{A_{\dots u_i v_j \dots}}{A_{\dots v_j u_i \dots}} &=
  S_{u_i v_j}
\end{alignat*}
with $S_{u_i v_j}$ given by~(\ref{equ:trimers:suv}).
Finally comes the product of all the $u_{\pi(i)}^{x_i}$
and~$v_{\sigma(j)}^{y_j}$.
The eigenvalue for the eigenfunction $\psi$ is given by
\begin{equation}
  \tilde\Lambda =
  \prod_{i=1}^{\nL} \frac{w_5}{w_0} u_i^{\frac{1}{2}}
  \prod_{j=1}^{\nR} \frac{w_1}{w_0} v_j^{-\frac{1}{2}}.
  \label{equ:trimers:eigenvalue}
\end{equation}
We have no rigorous proof that the above solution is correct for all
sectors, but using computer algebra we have verified it for
$\nL + \nR \le 5$.

It should be noted that the formulation of the solution depends on the
particle interpretation of the three-spin blocks.  The particle content
of each three-spin block is determined by
$\nL =
n_{\downarrow \bullet \bullet} + n_{\bullet \uparrow \bullet}$ and
$\nR =
n_{\bullet \bullet \downarrow} + n_{\bullet \uparrow \bullet}$, but the
order of the particles within a block can be chosen.  For example, we
could interpret $\downarrow\uparrow\uparrow$ as LLR, LRL or~RLL\@.  The
choices in Table~\ref{tab:trimers:blocks} lead to a simple description
of the eigenfunctions; each factor $D$ or $B$ depends only on two
successive particles.  Other choices than those in
Table~\ref{tab:trimers:blocks} would make the formulation of the
eigenfunctions more awkward; there would be more factors than just $D$
and $B$, and some would depend on non-successive particles.

\subsection{Bethe Ansatz equations}

Impose periodic boundary conditions in the horizontal direction.  This
means that the wave function must not change if the leftmost particle
(L at $x_1$ or R at $y_1$) is moved to the corresponding position
at the other (right) side of the system:
\begin{align*}
  \psi(\mathrm{L} \ x_1, \dots) &=
  \psi(\dots, \mathrm{L} \ x_1 + L), \\
  \psi(\mathrm{R} \ y_1, \dots) &=
  \psi(\dots, \mathrm{R} \ y_1 + L).
\end{align*}
The eigenfunction $\psi$ given by (\ref{equ:trimers:psigeneral})
satisfies these conditions if the Bethe Ansatz equations (BAEs) hold:
\begin{align}
  u_i^L &=
  (-)^{\nL-1} \prod_{j=1}^{\nR} S_{u_i v_j},
  \label{equ:trimers:baeus} \\
  v_j^L &=
  (-)^{\nR-1} \prod_{i=1}^{\nL} S_{u_i v_j}^{-1}.
  \label{equ:trimers:baevs}
\end{align}
Note that although the
description of an eigenfunction in terms of the $u$ and $v$ involves
factors (\ref{equ:trimers:b}) and (\ref{equ:trimers:d}), the BAEs only
contain factors~(\ref{equ:trimers:suv}).

Upon substitution of
\begin{equation}
  u = \left( \frac{w_0^3 w_3 w_4}{w_1 w_2 w_5^3}
      \right)^{\!\frac{1}{2}} \xi
  \qquad \mbox{and} \qquad
  v = -\left( \frac{w_1^3 w_4 w_5}{w_0^3 w_2 w_3}
       \right)^{\!\frac{1}{2}} \eta^{-1}
  \label{equ:trimers:substit}
\end{equation}
the expression (\ref{equ:trimers:suv}) for $S_{uv}$ simplifies to
\begin{displaymath}
  S_{uv} =
  \left( \frac{w_0 w_2 w_4}{w_1 w_3 w_5}
  \right)^{\!\frac{1}{2}}
  \frac{\eta - \xi}{1 + \xi \eta}.
\end{displaymath}
The BAEs (\ref{equ:trimers:baeus}) and (\ref{equ:trimers:baevs}) can
then be written
\begin{align}
  \left( \frac{w_0^3 w_3 w_4}{w_1 w_2 w_5^3}
  \right)^{\!\frac{1}{2} L}
  \left( \frac{w_1 w_3 w_5}{w_0 w_2 w_4}
  \right)^{\!\frac{1}{2} \nR} \xi_i^L &=
  (-)^{\nL + \nR - 1}
  \prod_{j=1}^{\nR}
  \eta_j^{-1} \frac{\xi_i - \eta_j}{\xi_i + \eta_j^{-1}},
  \label{equ:trimers:baeus2} \\
  \left( \frac{w_1^3 w_4 w_5}{w_0^3 w_2 w_3}
  \right)^{\!\frac{1}{2} L}
  \left( \frac{w_1 w_3 w_5}{w_0 w_2 w_4}
  \right)^{\!\frac{1}{2} \nL} \eta_j^L &=
  (-)^{L + \nR - 1} \prod_{i=1}^{\nL}
  \xi_i^{-1} \frac{\eta_j - \xi_i}{\eta_j + \xi_i^{-1}}.
  \label{equ:trimers:baevs2}
\end{align}
These equations can be considered the key result in the exact solution
of the model.  They determine the possible values for the $\xi$ and the
the~$\eta$.  These in turn determine the eigenvalues and the
eigenfunctions of the transfer matrix,
\begin{equation}
  \Lambda = w_0^L
  \left( \frac{w_3 w_4 w_5}{w_0 w_1 w_2}
  \right)^{\!\frac{1}{4} \nL}
  \left( \frac{w_1 w_2 w_3}{w_0 w_4 w_5}
  \right)^{\!\frac{1}{4} \nR}
  \biggl( \prod_{i=1}^{\nL} \xi_i
          \prod_{j=1}^{\nR} (-\eta_j)
  \biggr)^{\!\frac{1}{2}},
  \label{equ:trimers:eigenvalue2}
\end{equation}
where we have reintroduced the factor $w_0^L$ that was omitted as
of Subsection~\ref{sub:trimers:zerozero}.

As a check on the Bethe Ansatz we determined the eigenvalues of the
transfer matrix for small system size by (brute force) numerical
diagonalisation; the same eigenvalues were obtained by numerically
solving the~BAEs.

\section{Thermodynamics}
\label{sec:trimers:thermo}

We are interested in the behaviour of the model as a function of the
sub-lattice densities, that is, the canonical ensemble.  In order to
set up a transfer matrix we have passed to the grand canonical
ensemble, which is controlled by sub-lattice weights (or chemical
potentials) instead of sub-lattice densities.  In this section it turns
out that the transfer matrix leads
to a semi-grand canonical ensemble.  It is controlled partly by
densities (essentially the two conserved quantities) and partly by
chemical potentials.  We describe the Legendre transformation from this
ensemble back to the canonical ensemble.  We also look into the
symmetries between the sub-lattices and how they appear in the
semi-grand canonical ensemble.

\subsection{Legendre transformation}
\label{sub:trimers:legendre}

In passing to the grand canonical ensemble each trimer on a sub-lattice
$i$ was given a weight $w_i = \exp(\mu_i)$.  Certain combinations
of these weights occur in the BAEs (\ref{equ:trimers:baeus2}) and
(\ref{equ:trimers:baevs2}) and in the expression
(\ref{equ:trimers:eigenvalue2}) of the transfer matrix eigenvalue in
terms of the BA roots.  It is convenient to assign names to the
corresponding combinations of the chemical
potentials~$\mu_i = \log w_i$,
\begin{align*}
  \phiL &=
  \frac{1}{2} \bigl[
  \left( 3 \mu_0 - \mu_1 - \mu_2 + \mu_3 + \mu_4 - 3 \mu_5 \right) +
  \rhoR \left( - \mu_0 + \mu_1 - \mu_2 + \mu_3 - \mu_4 + \mu_5 \right)
  \bigr], \\
  \phiR &=
  \frac{1}{2} \bigl[
  \left( 3 \mu_0 - 3 \mu_1 + \mu_2 + \mu_3 - \mu_4 - \mu_5 \right) +
  \rhoL \left( - \mu_0 + \mu_1 - \mu_2 + \mu_3 - \mu_4 + \mu_5 \right)
  \bigr], \\
  \muL &=
  \frac{1}{4} \left( - \mu_0 - \mu_1 - \mu_2 + \mu_3 + \mu_4 + \mu_5
  \right), \\
  \muR &=
  \frac{1}{4} \left( - \mu_0 + \mu_1 + \mu_2 + \mu_3 - \mu_4 - \mu_5
  \right),
\end{align*}
where
\begin{displaymath}
  \rhoL = \frac{\nL}{L}
  \qquad \mbox{and} \qquad
  \rhoR = \frac{\nR}{L}
\end{displaymath}
denote the densities of the particles L and~R\@.
With these definitions
the BAEs (\ref{equ:trimers:baeus2}) and (\ref{equ:trimers:baevs2}) can
be written
\begin{align}
  \left( \mathrm{e}^{\phiL} \xi_i \right)^L &=
  (-)^{\nL + \nR - 1}
  \prod_{j=1}^{\nR}
  \eta_j^{-1} \frac{\xi_i - \eta_j}{\xi_i + \eta_j^{-1}},
  \label{equ:trimers:baexis} \\
  \left( \mathrm{e}^{\phiR} \eta_j \right)^L &=
  (-)^{L + \nR - 1} \prod_{i=1}^{\nL}
  \xi_i^{-1} \frac{\eta_j - \xi_i}{\eta_j + \xi_i^{-1}},
  \label{equ:trimers:baeetas}
\end{align}
while the eigenvalue expression (\ref{equ:trimers:eigenvalue2}) becomes
\begin{equation}
  \Lambda =
  \exp (L \mu_0 +
        \nL \muL +
        \nR \muR)
  \biggl( \prod_{i=1}^{\nL} \xi_i \;
          \prod_{j=1}^{\nR} (-\eta_j)
  \biggr)^{\frac{1}{2}}.
  \label{equ:trimers:eigenvalue3}
\end{equation}
Taking the logarithm, dividing by $L$, and letting $L$ to infinity
gives the free energy per trimer in the thermodynamic limit:
\begin{displaymath}
  \Omega(\rhoL, \rhoR;
                \mu_0, \mu_1, \dots, \mu_5) =
  \Phi(\rhoL, \rhoR;
              \phiL, \phiR) -
  \rhoL \muL -
  \rhoR \muR - \mu_0,
\end{displaymath}
where
\begin{equation}
  \Phi(\rhoL, \rhoR;
              \phiL, \phiR) =
  -\lim_{L \to \infty}
  \frac{1}{L} \log \biggl( \prod_{i=1}^{\nL} \xi_i \;
                           \prod_{j=1}^{\nR} (-\eta_j)
                   \biggr)^{\frac{1}{2}}.
  \label{equ:trimers:defPHI}
\end{equation}
It is the free energy in a semi-grand canonical ensemble where the
numbers of trimers on the different sub-lattices may vary but are
subject to the constraints imposed by fixing the particle densities
\begin{align}
  \rhoL &=
  1 - \rho_0 - \rho_1 + \rho_3 + \rho_4,
  \label{equ:trimers:rhol} \\
  \rhoR &=
  1 - \rho_0 + \rho_2 + \rho_3 - \rho_5.
  \label{equ:trimers:rhor}
\end{align}
In order to do the Legendre transform to the canonical ensemble the
derivatives of $\Omega$ with respect to $\mu_0$, $\mu_1$,
\dots, $\mu_5$ have to be taken.  This gives the ensemble average
densities
\begin{align}
  \rho_0 &=
  \left(- \frac{3}{2} + \frac{1}{2} \rhoR\right)
  \frac{\partial \Phi}{\partial \phiL} +
  \left(- \frac{3}{2} + \frac{1}{2} \rhoL\right)
  \frac{\partial \Phi}{\partial \phiR} -
  \frac{1}{4} \rhoL - \frac{1}{4} \rhoR +
  1.
  \label{equ:trimers:rho0} \\
  \rho_1 &=
  \left(+ \frac{1}{2} - \frac{1}{2} \rhoR\right)
  \frac{\partial \Phi}{\partial \phiL} +
  \left(+ \frac{3}{2} - \frac{1}{2} \rhoL\right)
  \frac{\partial \Phi}{\partial \phiR} -
  \frac{1}{4} \rhoL + \frac{1}{4} \rhoR,
  \\
  \rho_2 &=
  \left(+ \frac{1}{2} + \frac{1}{2} \rhoR\right)
  \frac{\partial \Phi}{\partial \phiL} +
  \left(- \frac{1}{2} + \frac{1}{2} \rhoL\right)
  \frac{\partial \Phi}{\partial \phiR} -
  \frac{1}{4} \rhoL + \frac{1}{4} \rhoR,
  \\
  \rho_3 &=
  \left(- \frac{1}{2} - \frac{1}{2} \rhoR\right)
  \frac{\partial \Phi}{\partial \phiL} +
  \left(- \frac{1}{2} - \frac{1}{2} \rhoL\right)
  \frac{\partial \Phi}{\partial \phiR} +
  \frac{1}{4} \rhoL + \frac{1}{4} \rhoR,
  \\
  \rho_4 &=
  \left(- \frac{1}{2} + \frac{1}{2} \rhoR\right)
  \frac{\partial \Phi}{\partial \phiL} +
  \left(+ \frac{1}{2} + \frac{1}{2} \rhoL\right)
  \frac{\partial \Phi}{\partial \phiR} +
  \frac{1}{4} \rhoL - \frac{1}{4} \rhoR,
  \\
  \rho_5 &=
  \left(+ \frac{3}{2} - \frac{1}{2} \rhoR\right)
  \frac{\partial \Phi}{\partial \phiL} +
  \left(+ \frac{1}{2} - \frac{1}{2} \rhoL\right)
  \frac{\partial \Phi}{\partial \phiR} +
  \frac{1}{4} \rhoL - \frac{1}{4} \rhoR.
  \label{equ:trimers:rho5}
\end{align}
In Subsection~\ref{sub:trimers:sublat} it was seen that because the
sub-lattice densities satisfy two constraints, four of them are
independent.  Equations
(\ref{equ:trimers:rho0})--(\ref{equ:trimers:rho5}) express the
sub-lattice densities in terms of only four quantities, namely
$\rhoL$, $\rhoR$,
$\frac{\partial \Phi}{\partial \phiL}$
and~$\frac{\partial \Phi}{\partial \phiR}$.  Therefore
these four quantities must be independent, and the sub-lattice
densities given by (\ref{equ:trimers:rho0})--(\ref{equ:trimers:rho5})
must satisfy the two constraints, (\ref{equ:trimers:linear})
and~(\ref{equ:trimers:quadratic}).  This can also be verified by direct
computation.
The entropy per
trimer is
\begin{equation}
  S(\rho_0, \rho_1, \dots, \rho_5) =
  - \Omega + \sum_{\ell=0}^5 \rho_{\ell} \mu_{\ell} =
  - \Phi +
  \frac{\partial \Phi}{\partial \phiL}
  \phiL +
  \frac{\partial \Phi}{\partial \phiR}
  \phiR.
  \label{equ:trimers:entropy}
\end{equation}
It is remarkable that the chemical potentials $\mu_0$,
$\muL$ and $\muR$ that occur in the expression
(\ref{equ:trimers:eigenvalue3}) for the eigenvalue have disappeared in
the Legendre transformation.  As a consequence $\Phi$ and hence the
densities $\rho_0$, $\rho_1$, \dots, $\rho_5$ and the entropy $S$ are
now functions of four parameters: the particle densities
$\rhoL$ and $\rhoR$ and the potential-like
quantities $\phiL$ and~$\phiR$.  These are just
the parameters that govern the BAEs (\ref{equ:trimers:baexis})
and~(\ref{equ:trimers:baeetas}).  This agrees with the fact that the
canonical ensemble also has four parameters.

\subsection{Symmetries of the parameter space}
\label{sub:trimers:paramsymm}

For the reference state of the BA sub-lattice 0 was chosen.  Since the
model is invariant under horizontal translations over a single lattice
edge sub-lattice~2 (or~4) could have been chosen instead.  The original
situation can be regained by renumbering the sub-lattices
$i \to i-2 (\mod 6)$.  The sub-lattice
densities $\rho'_i$ in the new numbering are related to the densities
$\rho_i$ in the old numbering by
\begin{equation}
  \rho'_0 = \rho_2, \quad
  \rho'_1 = \rho_3, \quad
  \mbox{etc.}
  \label{equ:trimers:transl}
\end{equation}
and analogously for the chemical potentials~$\mu_i$.  From this one
computes
\begin{displaymath}
  \rhoL' = 2-\rhoR,
  \qquad
  \rhoR' = 1+\rhoL-\rhoR,
  \qquad
  \phiL' = -\phiL+\phiR,
  \qquad
  \phiR' = -\phiL.
\end{displaymath}
Similarly the model is invariant under reflection in a horizontal line.
The corresponding sub-lattice renumbering is $i \to i+3 (\mod 6)$.
This gives:
\begin{displaymath}
  \rhoL' = 2-\rhoL,
  \qquad
  \rhoR' = 2-\rhoR,
  \qquad
  \phiL' = \phiL,
  \qquad
  \phiR' = \phiR.
\end{displaymath}
The model is also invariant under reflection in a vertical line.
For the line passing through sub-lattices 0 and 3 the renumbering is
$i \to -i (\mod 6)$.
Obviously this is nothing but interchanging
left and right, so
\begin{displaymath}
  \rhoL' = \rhoR,
  \qquad
  \rhoR' = \rhoL,
  \qquad
  \phiL' = \phiR,
  \qquad
  \phiR' = \phiL.
\end{displaymath}
Together these three transformations generate a group of order twelve.
In Subsection~\ref{sub:trimers:curves} we shall find four ``families''
of points
in the parameter space where the entropy of the model can be computed
exactly.  These four families turn out to be related by symmetries from
this group.
Note that
under this symmetry group the free energy $\Omega$ and the entropy
$S$ are invariant, so $\Phi$ transforms in a certain way.  For
example, for the translation (\ref{equ:trimers:transl}) the
transformation is
\begin{displaymath}
  \Phi(2 - \rhoR, 1 + \rhoL -
  \rhoR; -\phiL + \phiR,
  -\phiL) =
  \Phi(\rhoL, \rhoR; \phiL,
  \phiR) - \frac{1}{2} \phiL.
\end{displaymath}
Finally the model is invariant under some rotations.  As an example,
consider the rotation over $2\pi/3$ about an up triangle of the
lattice.  The sub-lattice renumbering is: $1 \to 3 \to 5$.  This does
not give a simple transformation of $\rhoL$,
$\rhoR$, $\phiL$ and~$\phiR$.  This
can be explained as follows.  In the definition of these four
parameters the direction in which the transfer matrix acts plays a
special role.  Rotations do not preserve this direction, in contrast to
the translation and the two reflections described above.
The symmetry group generated by all these operations is of order~36.

\section{Integral equations}
\label{sec:trimers:inteqs}

In Section~\ref{sec:trimers:betheansatz} two sets of BAEs were derived.
These equations can be solved numerically, for system size $L$ up to a
few hundred, say.  This can be done essentially in the full parameter
space.  (The regions where numerical complications arise can be mapped
to regions without such difficulties by means of symmetries from
Subsection~\ref{sub:trimers:paramsymm}.)
We however want to get analytic expressions for the
physical quantities of the model in thermodynamic limit.
In the present section the BAEs in the thermodynamic limit are turned
into two integral equations for two complex functions.  These functions
are multivalued, and their monodromy properties are obtained from the
integral equations.  The functions are then determined from their
monodromy and analyticity properties.  In the next section these
functions will be used to compute physical quantities of the model.

\subsection{Derivation}
\label{sub:trimers:inteqderiv}

We shall now in the usual fashion derive integral equations from the
BAEs (\ref{equ:trimers:baexis}) and~(\ref{equ:trimers:baeetas}).  The
logarithmic version of these BAEs is
\begin{alignat}{2}
  L \FL(\xi_i) &\equiv
  (\nL + \nR - 1) \pi \mathrm{i}
  && \quad (\mbox{mod $2 \pi \mathrm{i}$}),
  \label{equ:trimers:logbaexis} \\
  L \FR(\eta_j) &\equiv
  (L + \nR - 1) \pi \mathrm{i}
  && \quad (\mbox{mod $2 \pi \mathrm{i}$}),
  \label{equ:trimers:logbaeetas}
\end{alignat}
where
\begin{align}
  \FL(z) &=
  \log z -
  \frac{1}{L} \sum_{j=1}^{\nR}
  \left[ \log (z - \eta_j) - \log (z + \eta_j^{-1}) \right] +
  \phiL +
  \frac{1}{L} \sum_{j=1}^{\nR} \log \eta_j,
  \label{equ:trimers:fl} \\
  \FR(z) &=
  \log z -
  \frac{1}{L} \sum_{i=1}^{\nL}
  \left[ \log (z - \xi_i) - \log (z + \xi_i^{-1}) \right] +
  \phiR +
  \frac{1}{L} \sum_{i=1}^{\nL} \log \xi_i.
  \label{equ:trimers:fr}
\end{align}
The derivatives of these functions are
\begin{align}
  \fL(z) &=
  \frac{1}{z} -
  \frac{1}{L} \sum_{j=1}^{\nR}
  \biggl( \frac{1}{z - \eta_j} - \frac{1}{z + \eta_j^{-1}} \biggr),
  \label{equ:trimers:sumeqfl} \\
  \fR(z) &=
  \frac{1}{z} -
  \frac{1}{L} \sum_{i=1}^{\nL}
  \biggl( \frac{1}{z - \xi_i} - \frac{1}{z + \xi_i^{-1}} \biggr).
  \label{equ:trimers:sumeqfr}
\end{align}

\begin{figure}[!hbt]
  \psfrag{0}[tr][tr][1][0]{$0$}
  \psfrag{-2}[t][t][1][0]{$-2$}
  \psfrag{-1}[t][t][1][0]{$-1$}
  \psfrag{1}[t][t][1][0]{$1$}
  \psfrag{2}[t][t][1][0]{$2$}
  \psfrag{-2i}[Br][Br][1][0]{$-2\mathrm{i}$}
  \psfrag{-i}[Br][Br][1][0]{$-\mathrm{i}$}
  \psfrag{i}[Br][Br][1][0]{$\mathrm{i}$}
  \psfrag{2i}[Br][Br][1][0]{$2\mathrm{i}$}
  \hfil\includegraphics[scale=0.4]{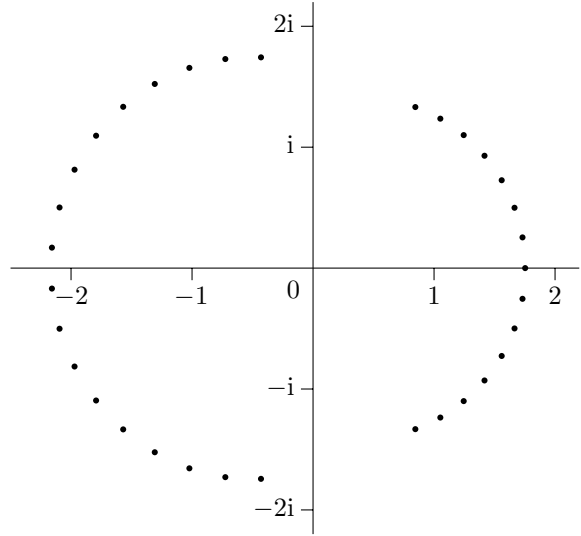}\hfil
  \caption{Distribution of the BAE roots for the largest eigenvalue.
    The $\xi$ are on the right, the $\eta$ on the left.  The
    parameters have the values $\phiL = -0.46$,
    $\phiR = -0.653$, $\nL = 15$,
    $\nR = 18$ and $L = 30$.}
  \label{fig:trimers:roots}
\end{figure}

For the understanding of the structure of the solutions to the BAEs we
rely on numeric computations for finite system size.  For many values
of the parameters $\rhoL$, $\rhoR$,
$\phiL$ and $\phiR$ the BA roots for the
largest eigenvalue show the following features.
The roots $\xi_i$ and $\eta_j$ lie on smooth curves in the
complex plane.  When the system size becomes large these curves tend to
well-defined limit shapes.  These limit curves will be called $\Xi$
and~$\Eta$.
The sets $\{\xi_i\}$ and $\{\eta_j\}$ (and hence also the curves
$\Xi$ and $\Eta$) are invariant under complex
conjugation; this implies that
\begin{equation}
  \fL(z^*) = \fL(z)^*
  \qquad \mbox{and} \qquad
  \fR(z^*) = \fR(z)^*.
  \label{equ:trimers:conjug}
\end{equation}
The curve $\Xi$ crosses the positive real axis,
whereas $\Eta$ crosses the negative real axis.
Note that by (\ref{equ:trimers:logbaexis}) the roots $\xi_i$ are
solutions of
\begin{equation}
  L \FL(\xi) \equiv (\nL + \nR - 1) \pi \mathrm{i}
  \quad (\mbox{mod $2 \pi \mathrm{i}$}).
  \label{equ:trimers:xidiscrete}
\end{equation}
This equation defines discrete points on the curve
$\Re \FL(\xi) = 0$.  The roots $\xi_i$ occupy a
succession of these points, without holes:
\begin{equation}
  L \left[ \FL(\xi_{i+1}) - \FL(\xi_i) \right] = 2 \pi \mathrm{i}.
  \label{equ:trimers:noholesxi}
\end{equation}
By holes we mean solutions of (\ref{equ:trimers:xidiscrete}) lying
between $\xi_1$ and $\xi_{\nL}$ on the curve
$\Re \FL(\xi) = 0$ that are not contained in the
set~$\{\xi_i\}$.  Similary for $\{\eta_j\}$:
\begin{equation}
  L \left[ \FR(\eta_{j+1}) - \FR(\eta_j) \right] = 2 \pi \mathrm{i}.
  \label{equ:trimers:noholeseta}
\end{equation}
Let $\bL$ and $\bR$ denote the end points
in the upper half plane of $\Xi$ and $\Eta$.
When the system becomes large $\xi_1$ and $\xi_{\nL}$ tend
to $\bL^*$ and $\bL$, respectively, while
$\eta_1$ and $\eta_{\nR}$ tend to $\bR$ and
$\bR^*$, respectively.
Figure~\ref{fig:trimers:roots} shows the distribution of the roots for
the largest eigenvalue in a given sector $\nL$,
$\nR$.

We assume that the condition that there are no holes in the sets of
roots $\{\xi_i\}$ and $\{\eta_j\}$ also holds in the
thermodynamic limit.  There (\ref{equ:trimers:noholesxi}) and
(\ref{equ:trimers:noholeseta}) can be written
\begin{align}
  L \fL(\xi_i) (\xi_{i+1} - \xi_i) &=
  2 \pi \mathrm{i},
  \label{equ:trimers:noholesxi2} \\
  L \fR(\eta_j) (\eta_{j+1} - \eta_j) &=
  2 \pi \mathrm{i},
  \label{equ:trimers:noholeseta2}
\end{align}
so that the sums in (\ref{equ:trimers:sumeqfl}) and
(\ref{equ:trimers:sumeqfr}) can be turned into integrals
\begin{align}
  \fL(z) &=
  \frac{1}{z} + \frac{1}{2 \pi \mathrm{i}} \int_{\Eta}
  \biggl( \frac{1}{\eta - z} + \frac{1}{\eta^{-1} + z} \biggr)
  \fR(\eta) \, \mathrm{d} \eta,
  \label{equ:trimers:inteqfl} \\
  \fR(z) &=
  \frac{1}{z} + \frac{1}{2 \pi \mathrm{i}} \int_{\Xi}
  \biggl( \frac{1}{\xi - z} + \frac{1}{\xi^{-1} + z} \biggr)
  \fL(\xi) \, \mathrm{d} \xi.
  \label{equ:trimers:inteqfr}
\end{align}
From these equations it is seen that $\fL(z)$ has branch
cuts $\Eta$ and $-\Eta^{-1}$ and that $\fR(z)$ has branch
cuts $\Xi$ and $-\Xi^{-1}$.  From the same equations it is easily
computed that $z \fL(z)$ and $z \fR(z)$ are
invariant under $z \mapsto -z^{-1}$.  Therefore we substitute
\begin{equation}
  z-z^{-1} = \hat z,
  \label{equ:trimers:ztohatz}
\end{equation}
and define
\begin{equation}
  z \fL(z) = \gL(\hat z)
  \qquad \mbox{and} \qquad
  z \fR(z) = \gR(\hat z).
  \label{equ:trimers:ftog}
\end{equation}
The two branch cuts $\Eta$ and $-\Eta^{-1}$ of $\fL(z)$ then
collapse to a single branch cut $\hat \Eta$ of $\gL(\hat z)$, and
similarly for~$\fR(z)$.  The equations
(\ref{equ:trimers:inteqfl}) and (\ref{equ:trimers:inteqfr}) become
\begin{align}
  \gL(\hat z) &=
  1 + \frac{1}{2 \pi \mathrm{i}}
  \int_{\hat \Eta} \frac{1}{\hat \eta - \hat z} \,
  \gR(\hat \eta) \, \mathrm{d} \hat \eta,
  \label{equ:trimers:inteqgl} \\
  \gR(\hat z) &=
  1 + \frac{1}{2 \pi \mathrm{i}}
  \int_{\hat \Xi} \frac{1}{\hat \xi - \hat z} \,
  \gL(\hat \xi) \, \mathrm{d} \hat \xi.
  \label{equ:trimers:inteqgr}
\end{align}
The functions $\fL(z)$ and $\fR(z)$ and
hence also $\gL(\hat z)$ and $\gR(\hat z)$
contain all information about the BA roots $\xi_i$ and $\eta_j$
that is needed to compute the densities $\rhoL$ and
$\rhoR$, the phases $\phiL$ and
$\phiR$, and the semi-grand canonical free
energy~$\Phi$.

The integral equations (\ref{equ:trimers:inteqgl}) and
(\ref{equ:trimers:inteqgr}) are very similar to the equations obtained
by Kalugin~\cite{kalugin:1994} for the square--triangle random tiling
model.  He tackles his equations by exploiting the monodromy properties
of the functions.  We shall use the same method for our integral
equations, closely following Kalugin's argument.

\subsection{Monodromy properties}

\begin{figure}[!hbt]
  \psfrag{0}[tr][tr][1][0]{$0$}
  \psfrag{-2}[t][t][1][0]{$-2$}
  \psfrag{-1}[t][t][1][0]{$-1$}
  \psfrag{1}[t][t][1][0]{$1$}
  \psfrag{-2i}[Br][Br][1][0]{$-2\mathrm{i}$}
  \psfrag{-i}[Br][Br][1][0]{$-\mathrm{i}$}
  \psfrag{i}[Br][Br][1][0]{$\mathrm{i}$}
  \psfrag{2i}[Br][Br][1][0]{$2\mathrm{i}$}
  \psfrag{Xh}[Bl][Bl][1][0]{$\hat \Xi$}
  \psfrag{Hh}[tr][tr][1][0]{$\hat \Eta$}
  \psfrag{GXh}[b][b][1][0]{$\Gamma_{\hat \Xi}$}
  \psfrag{GHh}[b][b][1][0]{$\Gamma_{\hat \Eta}$}
  \psfrag{GHhGXh}[b][b][1][0]{$\Gamma_{\hat \Eta} \Gamma_{\hat \Xi}$}
  \psfrag{(a)}[b][b][1][0]{(a)}
  \psfrag{(b)}[b][b][1][0]{(b)}
  \hfil\includegraphics[scale=0.4]{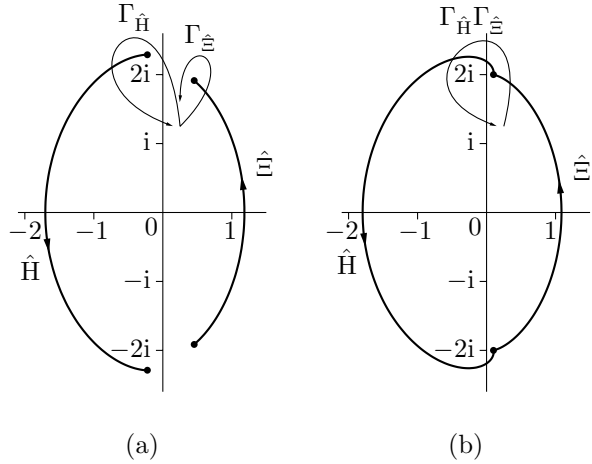}\hfil
  \caption{The contours $\hat \Xi$ and $\hat \Eta$.
    ($\phiL = -0.46$, $\phiR = -0.653$ and
    (a) $\rhoL = 0.5$, $\rhoR = 0.6$,
    (b) $\rhoL = 0.76$, $\rhoR = 0.93$.)}
  \label{fig:trimers:contours}
\end{figure}

The first step in solving the integral equations
(\ref{equ:trimers:inteqgl}) and (\ref{equ:trimers:inteqgr}) (but only
for a special case to be defined below) is to
establish the monodromy properties of the functions
$\gL(\hat z)$ and~$\gR(\hat z)$.  From
(\ref{equ:trimers:inteqgl}) it is seen that the contour
$\hat \Eta$ is a branch cut of the function
$\gL(\hat z)$.  Consider this function on a path
$\Gamma_{\hat \Eta}$ that crosses the contour
$\hat \Eta$ in some point $\hat z_0$, as in
Figure~\ref{fig:trimers:contours}(a).
The jump of the function $\gL$ over the contour is
\begin{displaymath}
  {\gL(\hat z_0)}_{\mathrm{after}} -
  {\gL(\hat z_0)}_{\mathrm{before}} =
  \frac{1}{2 \pi \mathrm{i}}
  \oint_{|\hat \eta - \hat z_0| = \varepsilon}
  \frac{1}{\hat \eta - \hat z_0} \,
  \gR(\hat \eta) \, \mathrm{d} \hat \eta =
  \gR(\hat z_0).
\end{displaymath}
Hence the analytic continuation of $\gL(\hat z)$ along
the path $\Gamma_{\hat \Eta}$ through the contour
$\hat \Eta$ is $\gL(\hat z) - \gR(\hat z)$.
From (\ref{equ:trimers:inteqgr}) the contour $\hat \Eta$ is not a
branch cut of the function $\gR(\hat z)$, so the
analytic continuation of $\gR(\hat z)$ along
$\Gamma_{\hat \Eta}$ is
just~$\gR(\hat z)$.
Therefore the effect of analytic continuation along
$\Gamma_{\hat \Eta}$ on a linear combination
$\aL \gL(\hat z) +
\aR \gR(\hat z)$ is given by
\begin{displaymath}
  \Gamma_{\hat \Eta} :
  \begin{pmatrix} \aL \\ \aR \end{pmatrix}
  \mapsto
  \begin{pmatrix} 1 & 0 \\ -1 & 1 \end{pmatrix}
  \begin{pmatrix} \aL \\ \aR \end{pmatrix}.
\end{displaymath}
It can be derived analogously that the monodromy operator for the path
$\Gamma_{\hat \Xi}$ in
Figure~\ref{fig:trimers:contours}(a) is given by
\begin{displaymath}
  \Gamma_{\hat \Xi} :
  \begin{pmatrix} \aL \\ \aR \end{pmatrix}
  \mapsto
  \begin{pmatrix} 1 & 1 \\ 0 & 1 \end{pmatrix}
  \begin{pmatrix} \aL \\ \aR \end{pmatrix}.
\end{displaymath}
The operators $\Gamma_{\hat \Eta}$ and
$\Gamma_{\hat \Xi}$ generate the full group
$\mathop{\rm SL}(2,\mathbf{Z})$.
Now consider the special case that the end points $\hat \bL$
and
$\hat \bL^*$ of $\hat \Xi$ coincide with the end
points $\hat \bR$ and $\hat \bR^*$ of
$\hat \Eta$, and that the contours do not meet in other points.
Then the paths $\Gamma_{\hat \Eta}$ and
$\Gamma_{\hat \Xi}$ are no longer defined, but their
composite $\Gamma_{\hat \Eta} \Gamma_{\hat \Xi}$
is, see Figure~\ref{fig:trimers:contours}(b).  Since this operator is
of
order six $\gL(\hat z)$ and $\gR(\hat z)$
are single-valued functions of the variable
\begin{equation}
  t = \left(
  \frac{\hat z - \hat b}{\hat z - \hat b^*} \right)^{\!\! 1/6},
  \label{equ:trimers:hatztot}
\end{equation}
where $\hat b = \hat \bL = \hat \bR$ and
$\hat b^* = \hat \bL^* = \hat \bR^*$ are the
common end points of the contours.  The inverse transformation is
\begin{displaymath}
  \hat z = \frac{\hat b^* t^6 - \hat b}{t^6 - 1}.
\end{displaymath}
Since $(\Gamma_{\hat \Eta} \Gamma_{\hat \Xi})
\gR(\hat z) = \gL(\hat z)$, the functions
$\gL(\hat z)$ and $\gR(\hat z)$ are
different branches of a single function~$g(\hat z)$.

In the remainder of this paper we shall, unless stated otherwise,
consider
the case that the end points of the contours
$\hat \Xi$ and $\hat \Eta$ coincide, and that these are the
only common points of these contours.

\subsection{Analyticity properties}

It follows from (\ref{equ:trimers:inteqgl}) that
$\gL(\hat z)$ is analytic everywhere except on the branch
cut~$\hat \Eta$.  Similarly $\gR(\hat z)$ is analytic
everywhere except on~$\hat \Xi$.  In particular it is analytic on
the contour~$\hat \Eta$, except perhaps at the end points, as these lie
also on~$\hat \Xi$.  It then follows from
(\ref{equ:trimers:inteqgl}) that $\gL(\hat z)$ remains
finite if $\hat z$ approaches a point (not an end point) on its branch
cut~$\hat \Eta$.  An analogous statement holds
for~$\gR(\hat z)$.  To summarise,
$\gL(\hat z)$ and $\gR(\hat z)$ are finite
everywhere except perhaps at $\hat b$ and~$\hat b^*$.

\begin{figure}[!hbt]
  \psfrag{t1}[bl][bl][1][0]{$t_1$}
  \psfrag{t2}[bl][bl][1][0]{$t_2$}
  \psfrag{t3}[bl][bl][1][0]{$t_3$}
  \psfrag{t4}[br][br][1][0]{$t_4$}
  \psfrag{t5}[br][br][1][0]{$t_5$}
  \psfrag{t6}[tr][tr][1][0]{$t_6$}
  \psfrag{g1}[B][B][1][0]{$\gL$}
  \psfrag{g1-2}[B][B][1][0]{$\gL - \gR$}
  \psfrag{g2}[B][B][1][0]{$-\gR$}
  \psfrag{g2-3}[B][B][1][0]{$-\gR$}
  \psfrag{g3}[B][B][1][0]{$-\gL - \gR$}
  \psfrag{g3-4}[B][B][1][0]{$-\gL$}
  \psfrag{g4}[B][B][1][0]{$-\gL$}
  \psfrag{g4-5}[B][B][1][0]{$-\gL + \gR$}
  \psfrag{g5}[B][B][1][0]{$\gR$}
  \psfrag{g5-6}[B][B][1][0]{$\gR$}
  \psfrag{g6}[B][B][1][0]{$\gL + \gR$}
  \psfrag{g6-1}[B][B][1][0]{$\gL$}
  \hfil\includegraphics[scale=0.4]{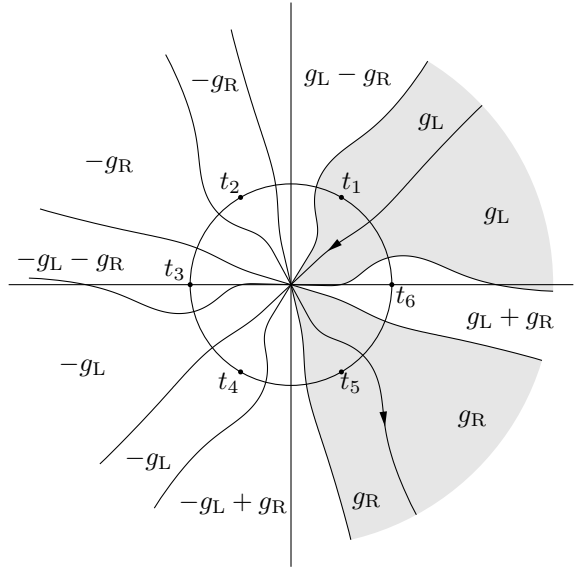}\hfil
  \caption{The complex $t$-plane.  The contours corresponding
    to $\hat \Xi$ and $\hat \Eta$ divide the plane into
    sectors that correspond to different branches of the
    function~$g(\hat z)$.  The shaded regions correspond to
    $\gL(\hat z)$ and~$\gR(\hat z)$.
    (The interest of this picture lies in its qualitative features, but
    it was actually computed from a numerical solution of the BAEs.
    The parameters are $\phiL = -0.46$,
    $\phiR = -0.653$, $\nL = 152$,
    $\nR = 186$, and $L = 200$.  These values correspond
    to~$\hat b = 2 \mathrm{i} \mathrm{e}^{-0.05 \mathrm{i}}$.)
    }
  \label{fig:trimers:tplane}
\end{figure}

It was derived above that they are branches of one function
$g(\hat z)$, which in turn is a single-valued function of~$t$.
Fix this function $h(t)$ by choosing that at
$t = \mathrm{e}^{\pi \mathrm{i}/3}$ it corresponds to
$\gL(\hat z)$ (at $\hat z = \infty$).
Figure~\ref{fig:trimers:tplane} shows where in the $t$-plane all the
branches of $g(\hat z) = h(t)$ are situated.  Note that the branch at
$t = \mathrm{e}^{-\pi \mathrm{i}/3}$ is $\gR(\hat z)$
(at~$\hat z = \infty$).

Since $g(\hat z)$ is finite everywhere except perhaps at
$\hat z = \hat b$ and $\hat z = \hat b^*$, $h(t)$ is analytic
everywhere except perhaps at $t = 0$ and~$t = \infty$.  Because $h(t)$
is single-valued it can only have power singularities (with integer
exponent).
Now
\begin{equation}
  \rhoL =
  \frac{1}{2 \pi} \int_{\Xi} \fL(z) \, \mathrm{d} z =
  \frac{1}{2 \pi} \int_{\hat \Xi} \gL(\hat z)
  \frac{\mathrm{d} z}{\mathrm{d} \hat z} \, \mathrm{d} \hat z =
  \frac{1}{2 \pi} \int_{\infty}^0 h(t)
  \frac{\mathrm{d} z}{\mathrm{d} \hat z}
  \frac{\mathrm{d} \hat z}{\mathrm{d} t} \, \mathrm{d} t,
  \label{equ:trimers:intrhol}
\end{equation}
where the last integral is over some contour running from $\infty$ to
0, is finite.  Since $\frac{\mathrm{d} z}{\mathrm{d} \hat z}$ remains
finite and non-zero for $t$ near 0 or $\infty$, and
\begin{displaymath}
  \frac{\mathrm{d} \hat z}{\mathrm{d} t} =
  \frac{6 (\hat b-\hat b^*) t^5}{(t^6-1)^2} \sim
  \begin{cases}
    t^5 & \mbox{if $t \to 0$}, \\
    t^{-7} & \mbox{if $t \to \infty$},
  \end{cases}
\end{displaymath}
it follows that $h(t)$ has at worst
singularities $t^{-5}$ at $t = 0$ and $t^5$ at $t = \infty$.
Hence,
the 1-form
\begin{displaymath}
  g(\hat z) \, \mathrm{d} \hat z =
  h(t) \frac{\mathrm{d} \hat z}{\mathrm{d} t} \, \mathrm{d} t
  \label{equ:trimers:oneform}
\end{displaymath}
is nonsingular at $t = 0$ and~$t = \infty$.

\subsection{Calculation of $g(\hat z)$}
\label{sub:trimers:calcg}

In the previous subsection it was shown that the 1-form
(\ref{equ:trimers:oneform})
is nonsingular at $t = 0$ and~$t = \infty$.  The only singularities it
can have are second order poles at the zeros $t_1$, $t_2$, \dots,
$t_6$ of $t^6-1$.  (These are the points in the $t$-plane
corresponding to $\hat z = \infty$.)  Therefore it can be written as
\begin{displaymath}
  g(\hat z) \, \mathrm{d} \hat z =
  h(t) \frac{\mathrm{d} \hat z}{\mathrm{d} t} \, \mathrm{d} t =
  \sum_{k=1}^6 \left\{ \frac{r_k}{t-t_k} +
  \frac{s_k}{(t-t_k)^2} \right\} \mathrm{d} t.
\end{displaymath}
The coefficients $r_k$ and $s_k$ are given by
\begin{displaymath}
  r_k =
  \Res_{t=t_k}
  h(t) \frac{\mathrm{d} \hat z}{\mathrm{d} t} \, \mathrm{d} t =
  \Res_{\hat z=\infty} g(\hat z) \mathrm{d} \hat z.
\end{displaymath}
and
\begin{displaymath}
  s_k =
  \Res_{t=t_k} (t-t_k)
  h(t) \frac{\mathrm{d} \hat z}{\mathrm{d} t} \, \mathrm{d} t =
  \bigl[ (t-t_k) \hat z \bigr]_{t=t_k}
  \Res_{\hat z=\infty}
  \hat z^{-1} g(\hat z) \mathrm{d} \hat z,
\end{displaymath}
where the appropriate branch of $g(\hat z)$ is to be taken.

The residues
$\Res_{\hat z=\infty}
g(\hat z) \mathrm{d} \hat z$ and
$\Res_{\hat z=\infty}
\hat z^{-1} g(\hat z) \mathrm{d} \hat z$ still have to be computed.
From (\ref{equ:trimers:inteqgl}) and
(\ref{equ:trimers:inteqgr}) one has
\begin{alignat*}{2}
  \Res_{\hat z=\infty}
  \gL(\hat z) \mathrm{d} \hat z &=
  -\frac{1}{2 \pi \mathrm{i}} \int_{\hat \Eta}
  \gR(\hat \eta) \, \mathrm{d} \hat \eta &&=:
  \RL, \\
  \Res_{\hat z=\infty}
  \gR(\hat z) \mathrm{d} \hat z &=
  -\frac{1}{2 \pi \mathrm{i}} \int_{\hat \Xi}
  \gL(\hat \xi) \, \mathrm{d} \hat \xi &&=:
  \RR
\end{alignat*}
and
\begin{align*}
  \Res_{\hat z=\infty}
  \hat z^{-1} \gL(\hat z) \mathrm{d} \hat z &= -1, \\
  \Res_{\hat z=\infty}
  \hat z^{-1} \gR(\hat z) \mathrm{d} \hat z &= -1.
\end{align*}
The residues for the other branches of $g(\hat z)$ follow by
application of the monodromy operators.
They are listed in Table~\ref{tab:trimers:polres}.
It follows from (\ref{equ:trimers:conjug}) that the integrals
$\RL$ and $\RR$ are real.

\begin{table}[!hbt]
  \caption{The poles and residues of $g(\hat z) \mathrm{d} \hat z$.}
  \label{tab:trimers:polres}
  \hfil
  \begin{tabular}{||c|c||c|c|c||}
  \hline
  $k$ & $t_k$ & $g$ &
  $\Res_{\hat z=\infty}
  g(\hat z) \mathrm{d} \hat z$ &
  $\Res_{\hat z=\infty}
  \hat z^{-1} g(\hat z) \mathrm{d} \hat z$ \\
  \hline
  1 & $\mathrm{e}^{\pi \mathrm{i}/3}$ & $\gL$
    & $\RL$ & $-1$ \\
  2 & $-\mathrm{e}^{-\pi \mathrm{i}/3}$ & $-\gR$
    & $-\RR$ & $1$ \\
  3 & $-1$ & $-\gL-\gR$
    & $-\RL-\RL$ & $2$ \\
  4 & $-\mathrm{e}^{\pi \mathrm{i}/3}$ & $-\gL$
    & $-\RL$ & $1$ \\
  5 & $\mathrm{e}^{-\pi \mathrm{i}/3}$ & $\gR$
    & $\RR$ & $-1$ \\
  6 & $1$ & $\gL+\gR$
    & $\RL+\RR$ & $-2$ \\
  \hline
  \end{tabular}
  \hfil
\end{table}
Combining these results gives after some algebra that
\begin{align}
  g(\hat z) &=
  \sum_{k=1}^6 \left\{ \frac{r_k}{t-t_k} + \frac{s_k}{(t-t_k)^2}
  \right\}
  \left( \frac{\mathrm{d} \hat z}{\mathrm{d} t} \right)^{\!-1}
  \nonumber \\
  &= (1-2C) t + (1-2C^*) t^{-1} + C (t+t^{-5}) + C^* (t^{-1} + t^5)
  \label{equ:trimers:g}
\end{align}
with
\begin{displaymath}
  C = \frac{1}{6} + \frac{1}{2 \sqrt{3} \Im \hat b}
  \left[ \mathrm{e}^{\pi \mathrm{i}/3} \RL -
         \mathrm{e}^{-\pi \mathrm{i}/3} \RR \right].
\end{displaymath}

We shall now argue in the generic case, $\hat b \ne 2 \mathrm{i}$,
that~$C = 0$.
From (\ref{equ:trimers:noholesxi2}) and (\ref{equ:trimers:noholeseta2})
the curves $\Xi$ and $\Eta$ are described by
$\Re [\fL(z) \, \mathrm{d} z] = 0$ and
$\Re [\fR(z) \, \mathrm{d} z] = 0$,
respectively, so the corresponding curves in the $t$-plane are both
solutions of
\begin{equation}
  \Re \left[ \frac{g(\hat z)}{z}
  \frac{\mathrm{d} z}{\mathrm{d} \hat z}
  \frac{\mathrm{d} \hat z}{\mathrm{d} t} \, \mathrm{d} t \right] = 0.
  \label{equ:trimers:tdiffeq}
\end{equation}
Note that $z$ and $\frac{\mathrm{d} z}{\mathrm{d} \hat z}$ are not
single-valued functions of $t$, but the two branches of
\begin{displaymath}
  \frac{1}{z} \frac{\mathrm{d} z}{\mathrm{d} \hat z} =
  \frac{1}{z+z^{-1}} = \frac{1}{\sqrt{\hat z^2+4}}
\end{displaymath}
differ only by a sign, which does not
influence~(\ref{equ:trimers:tdiffeq}).
The two different solutions of~(\ref{equ:trimers:tdiffeq})
corresponding to $\Xi$ and $\Eta$, respectively, meet at
$t = 0$ (and at $t = \infty$), so at these points the differential
equation
admits multiple solutions.  When $t \to 0$
\begin{displaymath}
  \frac{g(\hat z)}{z} \frac{\mathrm{d} z}{\mathrm{d} \hat z}
  \frac{\mathrm{d} \hat z}{\mathrm{d} t} \, \mathrm{d} t =
  6 \frac{\hat b - \hat b^*}{b+b^{-1}}
  \left[ C + (1-C)^* t^4 + O(t^6) \right] \mathrm{d} t,
\end{displaymath}
so this implies that~$C = 0$.

We shall now argue in the special case $\hat b = 2 \mathrm{i}$
that~$C = 0$.  When $t \to 0$
\begin{displaymath}
  f(z) \, \mathrm{d} z =
  \frac{g(\hat z)}{z} \frac{\mathrm{d} z}{\mathrm{d} \hat z}
  \frac{\mathrm{d} \hat z}{\mathrm{d} t} \, \mathrm{d} t =
  6 \left[ C t^{-3} + (1-C)^* t + O(t^3) \right] \mathrm{d} t
\end{displaymath}
(and similarly when $t \to \infty$).  The finiteness of the integral
(\ref{equ:trimers:intrhol}) (or its analogue for
$\rhoR$) implies that $C = 0$.

Now
(\ref{equ:trimers:g}) becomes
\begin{equation}
  g(\hat z) = t + t^{-1}.
  \label{equ:trimers:g2}
\end{equation}
The functions $\gL(\hat z)$ and
$\gR(\hat z)$ are obtained by taking the appropriate
branches $\tL(\hat z)$ and $\tR(\hat z)$
of~$t(\hat z)$.  The branch $\tL(\hat z)$ is determined
by $\tL(\infty) = \mathrm{e}^{\pi \mathrm{i}/3}$ and the
fact that it has $\hat \Eta$ as its only branch cut.  Similarly
$\tR(\hat z)$ is determined by
$\tR(\infty) = \mathrm{e}^{-\pi \mathrm{i}/3}$ and the
fact that it has $\hat \Xi$ as its only branch cut.

\section{Calculation of physical quantities}

In Section~\ref{sec:trimers:thermo} the relation was established
between the canonical ensemble we are interested in and the semi-grand
canonical ensemble that arises in the BA from
Section~\ref{sec:trimers:betheansatz}.  In
Section~\ref{sec:trimers:inteqs} BA information was encoded in two
complex functions satisfying a set of integral equations.  These
functions were then solved from those equations.  In the present
section the physical quantities occurring in
Section~\ref{sec:trimers:thermo} are extracted from the complex
functions determined in Section~\ref{sec:trimers:inteqs}.

\subsection{Calculation of $\rhoL$,
$\rhoR$, $\phiL$, $\phiR$,
and $\Phi$}
\label{sub:trimers:physquant1}

From (\ref{equ:trimers:ftog}) and (\ref{equ:trimers:g2})
$\fL(z)$ and $\fR(z)$ are both given by
\begin{equation}
  f(z) = \frac{t + t^{-1}}{z},
  \label{equ:trimers:f}
\end{equation}
with different branches of~$t$.  It was claimed in
Subsection~\ref{sub:trimers:inteqderiv} that the BA parameters
$\rhoL$, $\rhoR$, $\phiL$
and $\phiR$ and the semi-grand canonical free energy
$\Phi$ can be computed from the functions $\fL(z)$
and $\fR(z)$.
They depend on the point~$\hat b$.
The particle density $\rhoL$ was already computed
in~(\ref{equ:trimers:intrhol}):
\begin{equation}
  \rhoL =
  \frac{1}{2 \pi \mathrm{i}}
  \int_{\Xi} \fL(z) \, \mathrm{d} z.
  \label{equ:trimers:intrhol2}
\end{equation}
Because $\fL(z)$ is analytic this integral does not
depend on the precise shape of $\Xi$, but on its homology only.

Next $\phiL$ is calculated.
Since $\fL(z)$ is known the function
$\FL(z)$ is determined up to an integration constant.
The real part of this integration constant is fixed by
$\Re \FL(\bL) = 0$,
see~(\ref{equ:trimers:logbaexis}).  From
(\ref{equ:trimers:fl}) one has
\begin{displaymath}
  \Re \left[ \FL(z) + \FL(-z^{-1}) \right]
  = 2 \phiL.
\end{displaymath}
It is now easy to compute that
\begin{equation}
  \phiL =
  \frac{1}{2} \Re \int_{\bL}^{-\bL^{-1}} \!\!
  \fL(z) \, \mathrm{d} z.
  \label{equ:trimers:intphil}
\end{equation}

From (\ref{equ:trimers:defPHI}) the free energy $\Phi$ equals
$-(\SigmaL + \SigmaR)$ with
\begin{align*}
  \SigmaL =& \lim_{L \to \infty}
  \frac{1}{2} \frac{1}{L} \sum_{j=1}^{\nR} \log |\eta_j|, \\
  \SigmaR =& \lim_{L \to \infty}
  \frac{1}{2} \frac{1}{L} \sum_{i=1}^{\nL} \log |\xi_i|.
\end{align*}
Using (\ref{equ:trimers:fl}) one calculates
\begin{displaymath}
  \Re \Bigl. \bigl[ \FL(z) - \log z \bigr] \Bigr|_0^{\infty} =
  2 \, \frac{1}{L} \sum_{j=1}^{\nR} \log |\eta_j|,
\end{displaymath}
so
\begin{equation}
  \SigmaL =
  \frac{1}{4} \Re \int_0^{\infty}
  \left( \fL(z) - \frac{1}{z} \right) \, \mathrm{d} z.
  \label{equ:trimers:intSIGL}
\end{equation}
In (\ref{equ:trimers:intphil}) and (\ref{equ:trimers:intSIGL}) the
integral again only depends on the homology of the integration path,
not on its precise shape.  The real part of the integral even does not
depend at all on the path chosen between the integration end points,
but the imaginary part does.
This is because the indefinite integral (\ref{equ:trimers:fl}) is a sum
of logarithms with real prefactors, and distinct branches of a
logarithm differ by a multiple of $2 \pi \mathrm{i}$, which is purely
imaginary.

Replacing all subscripts L in (\ref{equ:trimers:intrhol2}),
(\ref{equ:trimers:intphil}) and (\ref{equ:trimers:intSIGL}) with R
yields
expressions of $\rhoR$, $\phiR$, and
$\SigmaR$ as integrals of functions involving
$\fR(z)$.
These integrals for $\rhoL$, $\rhoR$,
$\phiL$, $\phiR$,
$\SigmaL$ and $\SigmaR$ are of
the form $\int \! y \, \mathrm{d} z$, where the points $(y,z)$ lie on
an
algebraic curve of genus~5.  Therefore the indefinite integrals cannot
be expressed in terms of ``standard'' functions.  This does not prove
that the definite integrals we need cannot be expressed in terms of
standard functions, but it seems unlikely that it can be done.  Of
course they can be evaluated numerically.

\subsection{Calculation of
$\frac{\partial \Phi}{\partial \phiL}$ and
$\frac{\partial \Phi}{\partial \phiR}$}
\label{sub:trimers:physquant2}

The Legendre transformation in Subsection~\ref{sub:trimers:legendre}
involves the derivatives
$\frac{\partial \Phi}{\partial \phiL}$ and
$\frac{\partial \Phi}{\partial \phiR}$.
Unfortunately we have not been able to compute $\Phi$ as a
function of $\rhoL$, $\rhoR$,
$\phiL$, and $\phiR$ for all values of
these arguments.  Instead we have in
Subsection~\ref{sub:trimers:physquant1} computed
these parameters and the free energy in the case that the curves
$\hat \Xi$ and $\hat \Eta$ close, as a function of their
common end point
$\hat b = \hat \bL = \hat \bR$.
In order to still obtain the derivatives
$\frac{\partial \Phi}{\partial \phiL}$ and
$\frac{\partial \Phi}{\partial \phiR}$ we resort to
a perturbation analysis.  The details can be found in
Appendix~\ref{app:trimers:perturb}; here we only give some results.
An infinitesimally small complex parameter $C$ describes how far the
curves open up.  The thermodynamic parameters $\rhoL$,
$\rhoR$, $\phiL$, $\phiR$
and the free energy $\Phi$ then are
functions of $\hat b$ and~$C$.
If all their first-order partial derivatives are known,
$\frac{\partial \Phi}{\partial \phiL}$ and
$\frac{\partial \Phi}{\partial \phiR}$ can be
found
by applying the standard coordinate transformation formula to the
transformation between coordinates $\Re \hat b$, $\Im \hat b$, $\Re C$
and $\Im C$ on the one hand and $\rhoL$,
$\rhoR$, $\phiL$ and $\phiR$ on the
other hand.
The derivatives with respect to $\Re \hat b$ and $\Im \hat b$ can be
obtained immediately from the integral expressions in
Subsection~\ref{sub:trimers:physquant1}.  For the derivatives with
respect to $\Re C$ and $\Im C$ the perturbation analysis is needed.
It tells that to leading order in $C$ the parameters
$\rhoL$, $\rhoR$, $\phiL$,
$\phiR$ and the free energy
$\Phi = -(\SigmaL + \SigmaR)$
are again given by the integrals (\ref{equ:trimers:intrhol2}),
(\ref{equ:trimers:intphil}), (\ref{equ:trimers:intSIGL}) and their
analogues involving $\fR(z)$, where $f(z)$ is now given
by
\begin{equation}
  f(z) = \frac{t + t^{-1} + C \left(t^{-5} - t\right) +
  C^* \left(t^5 - t^{-1}\right)}{z}.
  \label{equ:trimers:fperturbed}
\end{equation}
This yields integral expressions for their derivatives with respect to
$\Re C$ and~$\Im C$.

The expressions thus obtained for the partial derivatives of
$\rhoL$, $\rhoR$, $\phiL$,
$\phiR$ and $\Phi$ with respect to $\Re \hat b$,
$\Im \hat b$, $\Re C$ and $\Im C$ were evaluated numerically for some
chosen value of~$\hat b$, and from this
$\frac{\partial \Phi}{\partial \phiL}$ and
$\frac{\partial \Phi}{\partial \phiR}$ were calculated.
These derivatives were also computed from numerical solutions of the
BAEs for large system size $L$ by numerical differentiation.
The results from
the two methods agree, which supports the perturbation analysis of
Appendix~\ref{app:trimers:perturb}.

\subsection{Configuration of $\Xi$ and $\Eta$}
\label{sub:trimers:curves}

\begin{figure}[!hbt]
  \psfrag{0}[tr][tr][1][0]{$0$}
  \psfrag{-2}[t][t][1][0]{$-2$}
  \psfrag{-1}[t][t][1][0]{$-1$}
  \psfrag{1}[t][t][1][0]{$1$}
  \psfrag{2}[t][t][1][0]{$2$}
  \psfrag{-2i}[Br][Br][1][0]{$-2\mathrm{i}$}
  \psfrag{-i}[Br][Br][1][0]{$-\mathrm{i}$}
  \psfrag{i}[Br][Br][1][0]{$\mathrm{i}$}
  \psfrag{2i}[Br][Br][1][0]{$2\mathrm{i}$}
  \psfrag{X}[Bl][Bl][1][0]{$\Xi$}
  \psfrag{H}[tr][tr][1][0]{$\Eta$}
  \psfrag{I}[Bl][Bl][1][0]{I}
  \psfrag{II}[Bl][Bl][1][0]{II}
  \psfrag{III}[Bl][Bl][1][0]{III}
  \psfrag{IV}[Bl][Bl][1][0]{IV}
  \hfil\includegraphics[scale=0.75]{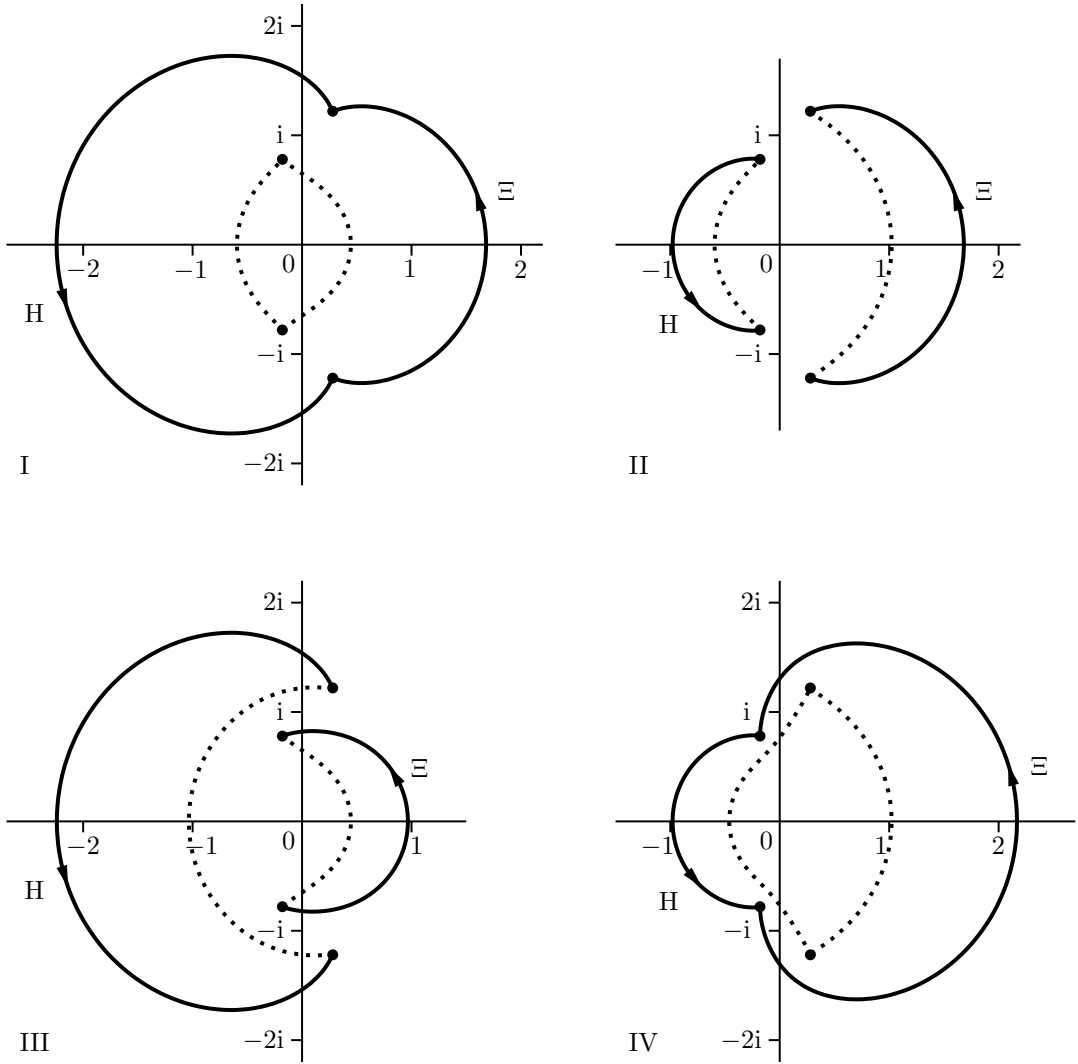}\hfil
  \caption{Four possible configurations of the curves $\Xi$ and~$\Eta$.
    The dashed curved are $-\Xi^{-1}$ and~$-\Eta^{-1}$.
    In cases I and IV $\Xi$ and $\Eta$ have the same end points.
    In cases II and III $\Xi$ and $-\Eta^{-1}$ share end points, as do
    $\Eta$ and $-\Xi^{-1}$.}
  \label{fig:trimers:zplanes}
\end{figure}

In the previous two sections
several physical quantities have been expressed as integrals of
functions involving $\fL(z)$ and $\fR(z)$.  These
integrals depend on the parameter $\hat b$ and on the topology of the
curves $\Xi$ and $\Eta$, but not on their precise shape.  If
$\hat b \ne 2 \mathrm{i}$ there are two distinct points in the
$z$-plane corresponding
to $\hat b$, say $b_1$ and $b_2$.  The end
points of $\Xi$ could be $b_1$ and $b_1^*$ or $b_2$ and
$b_2^*$, and the same holds for~$\Eta$.  Therefore one can expect
at least four different configurations for one and the same value
of~$\hat b$.
In order to determine what these four configurations actually are,
we first guessed what they might look like.  Then we
chose some particular value of $\hat b$ (close to $2 \mathrm{i}$) and
for each of the four expected cases computed the value of the
particle densities $\rhoL$ and $\rhoR$
and the phases $\phiL$ and~$\phiR$.  The
BAEs were solved numerically for these parameter values, for large
system size~$L$.  The resulting curves followed by the $\xi$ and the
$\eta$
display indeed the presupposed configurations.  These curves are shown
in Figure~\ref{fig:trimers:zplanes}.  Note that without first guessing
the configurations we would have had no way to find the values of the
parameters $\rhoL$, $\rhoR$,
$\phiL$ and $\phiR$, so there would have been
no BAEs to solve numerically.

The numerical results show that these four cases are related by the
symmetries of the parameter space discussed
in Subsection~\ref{sub:trimers:paramsymm}.  They are in a single orbit
of the
sub-group of order six generated by the horizontal translation (which
is of order three) and the product of the reflection in a horizontal
line and the reflection in a vertical line (which is of order two).
For the remaining two members of this orbit we have not been able to
numerically solve the BAEs.  In these cases the particle densities
are fairly high; we suppose that the curves $\Xi$ and $\Eta$
would cross or otherwise violate the condition that $\hat \Xi$
and $\hat \Eta$ only share their end points.

Once this symmetry is known a (numerical) calculation of the physical
quantities needs to be carried out only for one of the four
cases~I--IV\@.  The values for the other three cases are then obtained
at once by application of the symmetry transformations.

\subsection{Calculation of the sub-lattice densities and the entropy}
\label{sub:trimers:physquant3}

In Subsection~\ref{sub:trimers:physquant1} the physical quantities
$\rhoL$, $\rhoR$, $\phiL$,
$\phiR$, and $\Phi$ pertaining to the semi-grand
canonical ensemble were calculated from the functions
$\fL(z)$ and~$\fR(z)$.  In
Subsection~\ref{sub:trimers:physquant2} the derivatives
$\frac{\partial \Phi}{\partial \phiL}$ and
$\frac{\partial \Phi}{\partial \phiR}$ were
computed.  Substitution of these results into formulas
(\ref{equ:trimers:rho0})--(\ref{equ:trimers:rho5}) and
(\ref{equ:trimers:entropy}) from Subsection~\ref{sub:trimers:legendre}
gives the sub-lattice densities and the entropy, physical quantities
for the canonical ensemble.
This was performed numerically for a particular value of~$\hat b$.
The results reveal
that $\rho_0 = \rho_2 = \rho_4$ in cases I and IV and
$\rho_1 = \rho_3 = \rho_5$ in cases II and~III\@.
From the expressions
(\ref{equ:trimers:rho0})--(\ref{equ:trimers:rho5}) for the sub-lattice
densities this is equivalent to
\begin{equation}
  \frac{\partial \Phi}{\partial \phiL} =
  \frac{1}{6} (2 + \rhoL - 2 \rhoR)
  \qquad \mbox{and} \qquad
  \frac{\partial \Phi}{\partial \phiR} =
  \frac{1}{6} (2 - 2 \rhoL + \rhoR)
  \label{equ:trimers:derivsa}
\end{equation}
in cases I and IV and
\begin{equation}
  \frac{\partial \Phi}{\partial \phiL} =
  \frac{1}{6} (- \rhoL + 2 \rhoR)
  \qquad \mbox{and} \qquad
  \frac{\partial \Phi}{\partial \phiR} =
  \frac{1}{6} (2 \rhoL - \rhoR)
  \label{equ:trimers:derivsb}
\end{equation}
in cases II and~III\@.  One might hope to derive these expressions
analytically from the results of
Subsection~\ref{sub:trimers:physquant1}--\ref{sub:trimers:physquant2}.
We have not tried this because it would involve rather cumbersome
relations among integrals like (\ref{equ:trimers:intrhol2}),
(\ref{equ:trimers:intphil}) and~(\ref{equ:trimers:intSIGL}).
Once the expressions (\ref{equ:trimers:derivsa}) and
(\ref{equ:trimers:derivsb}) have been accepted
the perturbation analysis approach from
Subsection~\ref{sub:trimers:physquant2}
becomes superfluous.  Substituting them into
(\ref{equ:trimers:rho0})--(\ref{equ:trimers:rho5}) and
(\ref{equ:trimers:entropy}) yields new expressions for the sub-lattice
densities and the entropy.  The expressions for the sub-lattice
densities are polynomials in the particle densities
$\rhoL$ and $\rhoR$, the expression for
the entropy also contains the phases $\phiL$ and
$\phiR$ and of course the free energy~$\Phi$.

The cases I--IV correspond to different regions in the parameter space
of sub-lattice densities, as given in Table~\ref{tab:trimers:regions}.
These four cases are defined for $\Re \hat b > 0$ by
Figure~\ref{fig:trimers:zplanes}.  The mirror images (with respect to
the imaginary axis) of the configurations in
Figure~\ref{fig:trimers:zplanes} define cases I$'$--IV$'$ for
$\Re \hat b < 0$.  For example, the locus of $\Xi$ ($\Eta$) for case
I$'$ is the mirror image of the locus of $\Eta$ ($\Xi$) for case~I\@.
Table~\ref{tab:trimers:regions} also lists the regions in the parameter
space of sub-lattice densities corresponding to the cases I$'$--IV$'$.

\begin{table}[!hbt]
  \caption{The regions in parameter space of the sub-lattice densities
    for the cases I--IV and I$'$--IV$'$.}
  \label{tab:trimers:regions}
  \hfil
  \begin{tabular}{||c|c||c|c|c|c|c||}
    \hline
    \vrule width 0 pt height 2.5 ex
    $|\bL|$ & $|\bR|$ & & \multicolumn{2}{c|}{$\Re \hat b>0$} &
                          \multicolumn{2}{c||}{$\Re \hat b<0$} \\
    \hline
    $>1$ & $>1$   & $\rho_0 = \rho_2 = \rho_4$ &
           I      & $\rho_1 > \rho_5 > \rho_3$ &
           I$'$   & $\rho_5 > \rho_1 > \rho_3$ \\
    $>1$ & $<1$   & $\rho_1 = \rho_3 = \rho_5$ &
           II     & $\rho_0 > \rho_2 > \rho_4$ &
           III$'$ & $\rho_4 > \rho_0 > \rho_2$ \\
    $<1$ & $>1$   & $\rho_1 = \rho_3 = \rho_5$ &
           III    & $\rho_2 > \rho_0 > \rho_4$ &
           II$'$  & $\rho_0 > \rho_4 > \rho_2$ \\
    $<1$ & $<1$   & $\rho_0 = \rho_2 = \rho_4$ &
           IV     & $\rho_5 > \rho_3 > \rho_1$ &
           IV$'$  & $\rho_1 > \rho_3 > \rho_5$ \\
    \hline
  \end{tabular}
  \hfil
\end{table}

\subsection{Summary}
\label{sub:trimers:summary}

In the foregoing an exact solution of the trimer model was derived.
Because the results are obtained in the course of a long derivation,
we here provide a guide through them.  The final result is the entropy
as a function of six sub-lattice densities $\rho_i$ defined in
Subsection~\ref{sub:trimers:sublat}.  Complete coverage of the lattice
(\ref{equ:trimers:linear}) and a further geometric constraint
(\ref{equ:trimers:quadratic}) (derived in
Subsection~\ref{sub:trimers:constraint}) leave four independent
parameters.  A full analytic solution in the thermodynamic limit is
obtained in a two-parameter subspace.

The four-dimensional parameter space is described by new variables
$\rhoL$, $\rhoR$, $\phiL$ and~$\phiR$.  The free energy function of the
ensemble with these parameters is denoted by $\Phi$.
The sub-lattice densities are expressed in $\rhoL$, $\rhoR$,
$\frac{\partial \Phi}{\partial \phiL}$ and
$\frac{\partial \Phi}{\partial \phiR}$
in~(\ref{equ:trimers:rho0})--(\ref{equ:trimers:rho5}); the entropy is
given in terms of $\Phi$, $\phiL$, $\phiR$,
$\frac{\partial \Phi}{\partial \phiL}$ and
$\frac{\partial \Phi}{\partial \phiR}$ in~(\ref{equ:trimers:entropy}).

The free energy is written as a sum, $\Phi = -(\SigmaL + \SigmaR)$.
In the solvable subspace the quantities $\rhoL$, $\phiL$ and $\SigmaL$
are expressed as contour integrals of a function $\fL(z)$ in
(\ref{equ:trimers:intrhol2})--(\ref{equ:trimers:intSIGL}).  Analogously
the quantities $\rhoR$, $\phiR$ and $\SigmaR$ are integrals of a
function~$\fR(z)$.  The integration paths in the integral
(\ref{equ:trimers:intrhol2}) for $\rhoL$ and its analogue for $\rhoR$
are contours $\Xi$ and $\Eta$, respectively.  These contours are
symmetric under complex conjugation.  Their end points in the upper
half plane are denoted $\bL$ and $\bR$, respectively.  These satisfy
the equality $\bL - \bL^{-1} = \bR - \bR^{-1} = \hat b$.  For each
value of $\hat b$ this equation has two distinct solutions for $\bL$
and for $\bR$, resulting in four configurations I--IV of $\Xi$ and
$\Eta$ shown in Figure~\ref{fig:trimers:zplanes}.  The derivatives
$\frac{\partial \Phi}{\partial \phiL}$ and
$\frac{\partial \Phi}{\partial \phiR}$ are given by
(\ref{equ:trimers:derivsa}) in the cases I and IV and by
(\ref{equ:trimers:derivsb}) in the cases II and~III\@.

The functions $\fL(z)$ and $\fR(z)$ are different branches of a
function $f(z)$.  The branch cuts of $\fL(z)$ are $\Eta$ and
$-\Eta^{-1}$, and $\fR(z)$ has branch cuts $\Xi$ and $-\Xi^{-1}$.
In terms of a new variable $t$, defined by (\ref{equ:trimers:ztohatz})
and (\ref{equ:trimers:hatztot}), the function $f(z)$ is single-valued.
It is given by (\ref{equ:trimers:f}), while the functions $\fL(z)$ and
$\fR(z)$ are recovered by selecting the appropriate branch $\tL(z)$ and
$\tR(z)$ of $t$, specified at the end of
Subsection~\ref{sub:trimers:calcg}.

\section{Phase diagram}

In Subsections \ref{sub:trimers:sublat}
and~\ref{sub:trimers:constraint} a linear and a quadratic constraint on
the six sub-lattice densities were derived.  In this section we first
show that these constraints imply a breaking of the symmetry between
certain sub-lattices.  This symmetry breaking suggests that a phase
transition takes place when the total density
$\rhotd = \rho_1 + \rho_3 + \rho_5$ of down
trimers is increased from 0 to~1.
Next we compute the entropy as a function of $\rhotd$
from the exact solution of this model.  From this entropy the phase
diagram of the model in the parameter $\rhotd$ is
obtained.  It is also formulated in terms of the chemical potential of
the down trimers instead of their density.

\subsection{Symmetry breaking}

The linear constraint (\ref{equ:trimers:linear}) on the sub-lattice
densities can be rewritten as
\begin{equation}
  \rho_0 + \rho_2 + \rho_4 = 1 - \rhotd,
  \label{equ:trimers:linear2}
\end{equation}
and from the quadratic constraint (\ref{equ:trimers:quadratic}) one has
\begin{equation}
  \rho_0 \rho_2 + \rho_2 \rho_4 + \rho_4 \rho_0 \le
  \frac{1}{3} \rhotd^2
  \label{equ:trimers:quadratic2}
\end{equation}
with equality if and only if $\rho_1 = \rho_3 = \rho_5 =
\frac{1}{3} \rhotd$.  If $\rhotd$ is
small (to be precise: smaller than $2 \sqrt{3} - 3$), it follows from
(\ref{equ:trimers:linear2}) and (\ref{equ:trimers:quadratic2}) that one
of $\rho_0$, $\rho_2$ and $\rho_4$ is larger than the other
two, say $\rho_0 > \rho_2, \rho_4$.  Thus the symmetry between
the sub-lattices 0, 2 and 4 is broken.  If there is no further symmetry
breaking then $\rho_2 = \rho_4$ and
$\rho_1 = \rho_3 = \rho_5$, so
\begin{equation}
  \rho_0 > \rho_1 = \rho_3 = \rho_5 >
  \rho_2 = \rho_4.
  \label{equ:trimers:broken}
\end{equation}
By he same token the symmetry between the sub-lattices 1, 3 and 5 is
broken when $\rhotd$ is close to~1.  When
$\rhotd$ is increased from 0 to 1 the following seems to
be the simplest possible scenario.  At $\rhotd = 0$
sub-lattice 0 is fully occupied and the other sub-lattices are empty.
The six sub-lattice densities change continuously with
$\rhotd$, and (\ref{equ:trimers:broken}) holds up to
$\rhotd = \frac{1}{2}$.  There the six sub-lattice
densities are all equal to~$\frac{1}{6}$.  Then one of the odd
sub-lattices, say 3, takes over, and
\begin{displaymath}
  \rho_3 > \rho_0 = \rho_2 = \rho_4 >
  \rho_1 = \rho_5
\end{displaymath}
all the way to $\rhotd = 1$ where all trimers sit on
sub-lattice~3.

\subsection{Entropy for $\rhotd$}

In the previous subsection the occurrence was suggested of a phase
transition when $\rhotd$ is increased from 0 to~1.
For the study of such a  phase transition it would be helpful to know
the entropy of the model as a function
of~$\rhotd = \rho_1 + \rho_3 + \rho_5$.
However, what we have computed thus far is the entropy as a function of
all sub-lattice densities, but only for a two-dimensional subspace.
Therefore for given $\rhotd$ the sub-lattice densities
have to be determined for which the entropy is maximal.  If we are
fortunate these sub-lattice densities happen to lie in the
two-dimensional solved subspace.

For given $\rhotd < \frac{1}{2}$ the most symmetric
possibility for the six sub-lattice densities is described
by~(\ref{equ:trimers:broken}).  Another possibility,
\begin{equation}
  \rho_2 = \rho_4 > \rho_1 = \rho_3 = \rho_5 > \rho_0,
  \label{equ:trimers:broken2}
\end{equation}
exists when $2 \sqrt{3} - 3 \le \rhotd < \frac{1}{2}$.
Because of symmetry (\ref{equ:trimers:broken}) and
(\ref{equ:trimers:broken2}) are stationary points of the entropy.  It
is tempting to believe that (\ref{equ:trimers:broken}), being the more
general of the two most symmetric cases, corresponds to the maximum of
the entropy.  By numerically solving the BAEs the entropy of the model
can be computed to high precision.  Such calculations confirm
that for $\rhotd < \frac{1}{2}$ the entropy takes
its maximum at the symmetric case (\ref{equ:trimers:broken}) of the
sub-lattice densities, hence within the solvable subspace.

\begin{figure}[!hbt]
  \psfrag{0.00}[t][t][1][0]{$0.0$}
  \psfrag{0.25}[t][t][1][0]{$0.25$}
  \psfrag{0.50}[t][t][1][0]{$0.5$}
  \psfrag{0.75}[t][t][1][0]{$0.75$}
  \psfrag{1.00}[t][t][1][0]{$1.0$}
  \psfrag{0.0}[Br][Br][1][0]{$0.0$}
  \psfrag{0.1}[Br][Br][1][0]{$0.1$}
  \psfrag{0.2}[Br][Br][1][0]{$0.2$}
  \psfrag{0.3}[Br][Br][1][0]{$0.3$}
  \psfrag{S}[Br][Br][1][0]{$S$}
  \psfrag{rhodown}[t][t][1][0]{$\rhotd$}
  \hfil\includegraphics[scale=0.4]{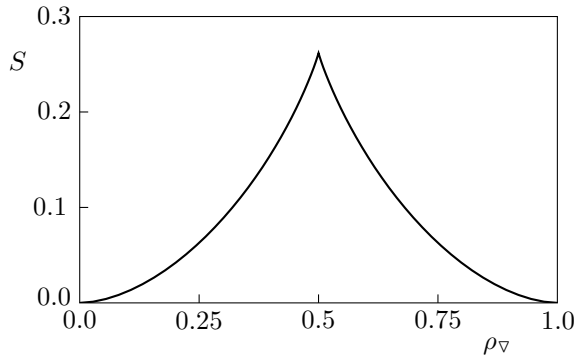}\hfil
  \caption{The entropy per trimer $S$ as a function of the total
    density of down trimers
    $\rhotd = \rho_1 + \rho_3 + \rho_5$.  It is obtained
    from the exact solution in the special case
    (\protect\ref{equ:trimers:broken}) for
    $\rhotd \le \frac{1}{2}$, and similarly for
    $\rhotd \ge \frac{1}{2}$.}
  \label{fig:trimers:entropy}
\end{figure}

\begin{figure}[!hbt]
  \psfrag{0.10}[t][t][1][0]{$0.10$}
  \psfrag{0.15}[t][t][1][0]{$0.15$}
  \psfrag{0.20}[t][t][1][0]{$0.20$}
  \psfrag{0.25}[t][t][1][0]{$0.25$}
  \psfrag{-0.3}[Br][Br][1][0]{$-0.3$}
  \psfrag{-0.2}[Br][Br][1][0]{$-0.2$}
  \psfrag{-0.1}[Br][Br][1][0]{$-0.1$}
  \psfrag{0.0}[Br][Br][1][0]{$0.0$}
  \psfrag{0.1}[Br][Br][1][0]{$0.1$}
  \psfrag{0.2}[Br][Br][1][0]{$0.2$}
  \psfrag{0.3}[Br][Br][1][0]{$0.3$}
  \psfrag{S}[Br][Br][1][0]{$S$}
  \psfrag{rhodiff}[t][t][1][0]{$\rho_2-\rho_4$}
  \hfil\includegraphics[scale=0.4]{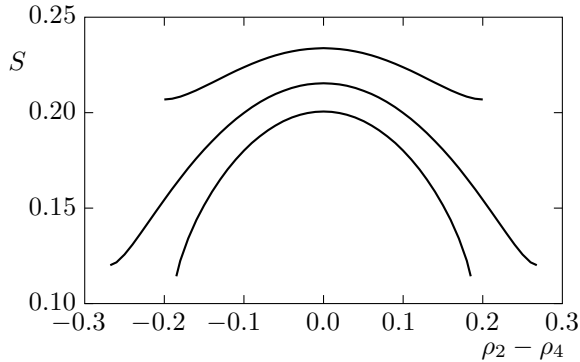}\hfil
  \caption{The entropy per trimer $S$ as a function of $\rho_2-\rho_4$
    at fixed $\rho_1 = \rho_3 = \rho_5$.  The end points of the lower
    curve ($\rhotd = 0.45$) are determined by
    $\rho_2 = 0$ and by~$\rho_4 = 0$.  The upper curve
    ($\rhotd = 0.48$) terminates when $\rho_4 = \rho_0$
    and when~$\rho_2 = \rho_0$.  At the left end point of the middle
    curve ($\rhotd = 2 \sqrt{3} - 3 \approx 0.4641$)
    $\rho_2 = 0$ and $\rho_4 = \rho_0$, at the right end point
    $\rho_4 = 0$ and~$\rho_2 = \rho_0$.}
  \label{fig:trimers:nufig}
\end{figure}

As was seen in Subsection~\ref{sub:trimers:physquant3}, for the
solvable subspace one has $\rho_0 = \rho_2 = \rho_4$ in cases
I and IV and $\rho_1 = \rho_3 = \rho_5$ in cases II and~III\@.
Consider case II and take $\hat b$ on the imaginary axis between 0
and~$2 \mathrm{i}$.  The contours $\hat \Xi$ and $\hat \Eta$ then
lie symmetric with respect to the imaginary axis, so
$\rhoL = \rhoR$, and hence
$\rho_2 = \rho_4$.  Thus this is precisely the symmetric
case~(\ref{equ:trimers:broken}).  Therefore we have obtained the
entropy as a function of $\rhotd$ for
$\rhotd < \frac{1}{2}$.  The entropy for
$\rhotd > \frac{1}{2}$ follows immediately by the
symmetry between up and down trimers.  This entropy can also be
obtained by
considering case I and taking $\hat b$ above $2 \mathrm{i}$ on the
imaginary axis.  The resulting entropy is shown in
Figure~\ref{fig:trimers:entropy}.
When in case II $\hat b$ is not taken on the imaginary axis between 0
and~$2 \mathrm{i}$, $\rho_2 \ne \rho_4$.
Figure~\ref{fig:trimers:nufig} shows the entropy as a function of the
asymmetry $\rho_2 - \rho_4$ at fixed $\rho_1 = \rho_3 = \rho_5$ along
the line determined by the constraints (\ref{equ:trimers:linear})
and~(\ref{equ:trimers:quadratic}).

For $\hat b = 2 \mathrm{i}$ all four cases I, II, III and IV coincide.
The integrals in Subsection~\ref{sub:trimers:physquant1} then simplify.
The sub-lattice densities are all equal to $\frac{1}{6}$, and the
entropy per trimer is
$S_{\mathrm{sym}} = \log \frac{4}{3} \sqrt{3}$.

\subsection{Phase transition}

Consider a system with $\rhotd$ between 0
and~$\frac{1}{2}$.  The energy is a convex function of
$\rhotd$, so the system is thermodynamically unstable.
It would separate into a frozen phase with
$\rhotd = 0$ and the symmetric phase with
$\rhotd = \frac{1}{2}$.  However an
interface between these two phases is not possible in the model.
Similarly a system with $\rhotd$ between $\frac{1}{2}$
and $1$ would demix into phases with
$\rhotd = \frac{1}{2}$ and
$\rhotd = 1$, if coexistence between these phases were possible.

Now give a chemical potential $\mu_{\triangledown}$ to the down
trimers instead of imposing their density~$\rhotd$.
The free energy
\begin{displaymath}
  F = -\mu_{\triangledown} \rhotd -
  S(\rhotd)
\end{displaymath}
takes its global minimum at
\begin{displaymath}
  \rhotd =
  \begin{cases}
    0           & \mbox{for $\phantom{-2 S_{\mathrm{sym}} \le {}}
                  \mu_{\triangledown} \le -2 S_{\mathrm{sym}}$} \\
    \frac{1}{2} & \mbox{for $-2 S_{\mathrm{sym}} \le \mu_{\triangledown}
                  \le \phantom{-} 2 S_{\mathrm{sym}}$} \\
    1           & \mbox{for $\phantom{-} 2 S_{\mathrm{sym}} \le
                  \mu_{\triangledown}$}
  \end{cases}
\end{displaymath}
Therefore the model is in a frozen phase for
$\mu_{\triangledown} < -2 S_{\mathrm{sym}}$ and for
$\mu_{\triangledown} > 2 S_{\mathrm{sym}}$ and in the symmetric
phase for $-2 S_{\mathrm{sym}} < \mu_{\triangledown} <
2 S_{\mathrm{sym}}$.
At $\mu_{\triangledown} = -2 S_{\mathrm{sym}}$ and
at $\mu_{\triangledown} = 2 S_{\mathrm{sym}}$ there is
coexistence between a frozen and a symmetric phase.

\section{Conclusion}

We have introduced a new simple lattice model.  It is a fluid of
particles each occupying three sites of the triangular lattice.  We
distinguish six sub-lattices of adsorption sites for the trimers.  Full
occupancy and a resulting geometric constraint leave of the six
sub-lattice densities four independent parameters.

In the full four-dimensional parameter space the model is solvable by
the Bethe Ansatz.  In the thermodynamic limit the Bethe Ansatz
equations can be reduced to two integral equations.  In a
two-dimensional subspace of the sub-lattice densities these integral
equations can be solved by means of monodromy and analyticity
properties of the functions involved.  Within this subspace the entropy
and the sub-lattice densities are given as integral expressions.

The solution is
very similar to that of the square--triangle random tiling
model~\cite{widom:1993,kalugin:1994}.  In
both cases the solution is closely connected to the hexagonal domain
wall structure of the model.  Another solvable model with such a domain
wall structure is the three-colouring model on the honeycomb
lattice~\cite{baxter:1970}.
In a configuration of this model the edges of the honeycomb lattice are
coloured with three colours in such a way that the three edges meeting
in each vertex have different colours.  Alternatively this model can be
formulated as the zero-temperature antiferromagnetic three-state Potts
model on the Kagom\'e lattice~\cite{wu:1982,huse:1992}.
We shall now briefly discuss the
relation between these three models.

\begin{figure}[!hbt]
  \hfil\includegraphics[scale=0.4]{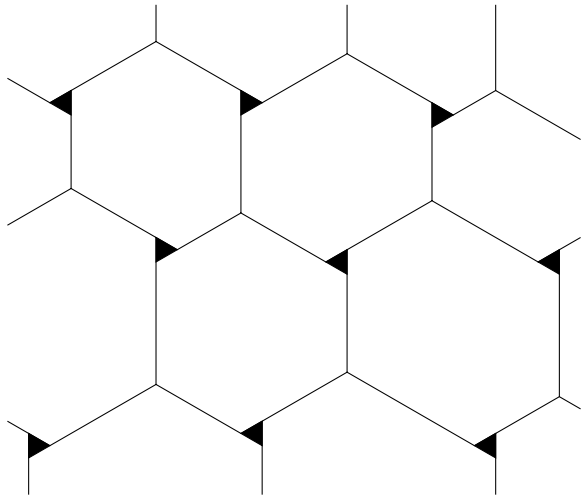}\hfil
  \caption{Schematic representation of the domain wall structure of the
    trimer model.  The Y-joints of the domain walls come in two types.
    These are mirror images, featuring either a $\blacktriangleleft$ or
    a~$\blacktriangleright$.  In contrast there is only one type of
    upside-down-Y-joints.}
  \label{fig:trimers:network}
\end{figure}

The domain wall structure of the trimer model is depicted schematically
in Figure~\ref{fig:trimers:network}.  It contains two types of Y-joints
but only one type of upside-down-Y-joints.  In the square--triangle
tiling there
is only one type of Y-joints and one type of upside-down-Y-joints.  In
the
honeycomb lattice three-colouring model on the other hand both the
Y-joints and the upside-down-Y-joints come in two types.  Hence these
three
models appear to be different.

The $A_2^{(1)}$ model is a vertex model on the square lattice, derived
from an affine Lie algebra~\cite{bazhanov:1985,jimbo:1986}.  It
satisfies the Yang--Baxter equation~\cite{baxter:1972,baxter:1982}, so
it can be solved by algebraic Bethe Ansatz~\cite{faddeev:1984}.  At a
special value of the spectral parameter it is the three-colouring model
on the honeycomb-lattice~\cite{reshetikhin:1991}.  For a suitable
choice of the remaining parameters one of the two types of Y-joints and
upside-down-Y-joints in the domain wall network is excluded.  In this
limit the
model is just the square--triangle tiling.  This mapping ``explains''
the solvability of the square--triangle tiling in terms of that of the
$A_2^{(1)}$ model~\cite{gier:1997b}.

In a similar fashion the square--triangle tiling can also be obtained
from the trimer model.  When the trimers on sub-lattice 4 (or 2) are
excluded, one (or the other) type of Y-joint no longer occurs in the
domain wall network.  Again the square--triangle tiling results.
The Bethe Ansatz for the trimer model remains valid in this special
case.  However, the substitutions (\ref{equ:trimers:substit}) no longer
makes sense when $w_4 = 0$ (or $w_2 = 0$), so the same is true of the
analysis in the subsequent sections.

Therefore the three models are connected in sense that both the trimer
model and the $A_2^{(1)}$ model contain the square--triangle random
tiling as a singular limit.  It would be interesting to know if the
trimer model, like the square--triangle tiling, is a special case of
some model satisfying the Yang--Baxter equation.

\section*{Acknowledgement}

We are grateful to Jan de Gier for many useful discussions, as well as
for putting at our disposal his computer program for numerical Bethe
Ansatz calculations.
We thank Dave Rusin for enlightening us on the question whether the
integrals from Subsection~\ref{sub:trimers:physquant1} can be expressed
in closed form.
This work is part of the research programme of the ``Stich\-ting voor
Fun\-da\-men\-teel On\-der\-zoek der Ma\-te\-rie (FOM)'', which is
financially supported by the ``Ne\-der\-land\-se Or\-ga\-ni\-sa\-tie
voor We\-ten\-schap\-pe\-lijk On\-der\-zoek (NWO)''.

\appendix

\section{Perturbation analysis}
\label{app:trimers:perturb}

In Subsection~\ref{sub:trimers:physquant1} the quantities
$\rhoL$, $\rhoR$, $\phiL$,
$\phiR$ and
$\Phi = -(\SigmaL + \SigmaR)$ were obtained as
functions of $\hat b = \hat \bL = \hat \bR$.
For the computation of the sub-lattice densities and the entropy the
derivatives
\begin{displaymath}
  \left(\frac{\partial \Phi}{\partial \phiL}\right)_
  {\rhoL \rhoR \phiR}
  \qquad \mbox{and} \qquad
  \left(\frac{\partial \Phi}{\partial \phiR}\right)_
  {\rhoL \rhoR \phiL}
\end{displaymath}
are also needed
as functions
of~$\hat b$.
These cannot be calculated by differentiation of the $\Phi$
already obtained, because variation of $\phiL$
($\phiR$) at constant $\rhoL$,
$\rhoR$ and $\phiR$ ($\phiL$)
breaks the condition $\hat \bL = \hat \bR$.
Therefore in this appendix we infinitesimally relax that condition, and
compute $\rhoL$, $\rhoR$, $\phiL$,
$\phiR$ and
$\Phi = -(\SigmaL + \SigmaR)$ to leading order
in the infinitesimal relaxation parameter.

When the curves $\hat \Xi$ and $\hat \Eta$ do not
close, the monodromy group is the full group
$\mathop{\rm SL}(2,\mathbf{Z})$.  Therefore $\gL(\hat z)$
and $\gR(\hat z)$ are no longer single-valued functions
of the variable~$t$.
Kalugin~\cite{kalugin:1994} has provided a perturbation analysis for
the analogous situation in the square--triangle random tiling model.
It leans heavily on the understanding of the structure of the Riemann
surface of the functions.  Our approach does not require such
knowledge and is more systematic.

Although $\gL(\hat z)$ and $\gR(\hat z)$ are no
longer single-valued functions of the variable~$t$,
one can still perform the variable
transformation~(\ref{equ:trimers:hatztot}).  The end points
$\hat \bL$ and $\hat \bL^*$ of
$\hat \Xi$ ($\hat \bR$ and
$\hat \bR^*$ of $\hat \Eta$) then correspond to
points $\dL$ and $\dL^{*-1}$
($\dR$ and $\dR^{*-1}$) in the $t$-plane.  The
point $\hat b$ in (\ref{equ:trimers:hatztot}) can be chosen such that
$|\dL| = |\dR|$; write
\begin{displaymath}
  \dL = \betaL \delta
  \qquad \mbox{and} \qquad
  \dR = \betaR \delta,
\end{displaymath}
with $\delta$ real and positive and
$|\betaL| = |\betaR| = 1$.
The curves corresponding to $\hat \Xi$ and $\hat \Eta$
divide the annulus $\delta < |t| < \delta^{-1}$ into sectors, much as
in Figure~\ref{fig:trimers:tplane}.  We get a single-valued function
$h(t) = g(\hat z)$ in this annulus instead of in the whole $t$-plane.
Since it is analytic in the annulus it can be expanded as a Laurent
series in~$t$:
\begin{equation}
  g(\hat z) = h(t) = \sum_{p=-\infty}^{\infty} A_p t^p.
  \label{equ:trimers:laurent}
\end{equation}
Only powers $t^p$ with $p \equiv \pm 1 \; (\mbox{mod 6})$ have the
correct monodromy properties, so other powers cannot occur.
From (\ref{equ:trimers:conjug}) one has
\begin{displaymath}
  h(t^{*-1}) = g(\hat z^*) = g(\hat z)^* = h(t)^*,
\end{displaymath}
so the coefficients $A_p$ satisfy
\begin{equation}
  A_{-p} = A_p^*\,.
  \label{equ:trimers:posnegsymm}
\end{equation}
We want to view the function $g(\hat z)$ given by
(\ref{equ:trimers:laurent}) as a perturbation of the function
$g(\hat z)$ given by~(\ref{equ:trimers:g2}), where $\delta$ is the
small parameter.  In our notation we have suppressed the dependence of
the coefficients $A_p$ on $\delta$, $\betaL$, and
$\betaR$.

The function $g(\hat z)$ satisfies the integral
equation~(\ref{equ:trimers:inteqgl}).
We investigate how each of the terms $t^p$ from the Laurent series
(\ref{equ:trimers:laurent}) of $g(\hat z)$ behaves in this equation.
In order to compute the integral we change from $\hat \eta$ to
$\tau=t(\hat \eta)$ as integration variable.  The resulting integrand
is a rational function in $\tau$, which we decompose into partial
fractions.  Integration yields polynomial as well as logarithmic terms;
some care is required in choosing the branch of the logarithms.
Finally we expand in powers of $\delta$, obtaining
\begin{align}
  \frac{1}{2 \pi \mathrm{i}} \int_{\hat \Eta}
  \frac{1}{\hat \eta - \hat z} \, \tau^p \, \mathrm{d} \hat \eta
  &=
  t^p - t_1^p + \frac{6}{2 \pi \mathrm{i}} \, \Biggl\{
  \sum_{q=-\infty}^{-1} \frac{\betaR^{p-6q}}{p-6q} \,
  (t^{6q} - 1) \, \delta^{p-6q} \; + \nonumber \\
  & \phantom{= t^p - t_1^p + \frac{6}{2 \pi \mathrm{i}} \, \Biggl(\{}
  \sum_{q=1}^{\infty} \, \frac{\betaR^{p-6q}}{p-6q} \,
  (t^{6q} - 1) \, \delta^{6q-p} \Biggr\}
  \label{equ:trimers:laurentintl}
\end{align}
for each term $t^p$ in the Laurent series~(\ref{equ:trimers:laurent}).
Here $t$ in the RHS corresponding to $\hat z$ in the LHS is in the
sector containing $t_1$, that is the sector where $g(\hat z)$ equals
$\gL(\hat z)$.  Comparison with the integral equation
(\ref{equ:trimers:inteqgl})
shows the following.  The term $t^p$ in the RHS of
(\ref{equ:trimers:laurentintl}) exactly matches the term
$\gL(\hat z)$ in the LHS of~(\ref{equ:trimers:inteqgl}).
The inhomogeneous term $-t_1^p$ in the RHS of
(\ref{equ:trimers:laurentintl}) corresponds to
the inhomogeneous term $1$ in the
integral equation.  The other terms in the RHS
of (\ref{equ:trimers:laurentintl}) are ``unwanted''; the powers of $t$
they involve are multiples of~6.  Because the Laurent series
(\ref{equ:trimers:laurent}) satisfies the integral equation
(\ref{equ:trimers:inteqgl}) the inhomogeneous terms $-t_1^p$ from
(\ref{equ:trimers:laurentintl}) counterbalance the
inhomogeneous term $1$ of the integral equation,
\begin{equation}
  \sum_{p=-\infty}^{\infty} t_1^p A_p = 1
  \label{equ:trimers:one}
\end{equation}
(which means that $\gL(\infty) = 1$), and the unwanted
terms cancel,
\begin{align}
  \sum_{p=-\infty}^{\infty} \frac{\betaR^p}{p-6q} \,
  \delta^p A_p
  &= 0 \qquad \mbox{for all $q<0$},
  \label{equ:trimers:cancelnegq} \\
  \sum_{p=-\infty}^{\infty} \frac{\betaR^p}{p-6q} \,
  \delta^{-p} A_p
  &= 0 \qquad \mbox{for all $q>0$}.
  \label{equ:trimers:cancelposq}
\end{align}
(Due to~(\ref{equ:trimers:posnegsymm}) the equations for $q$ and for
$-q$ are equivalent.)  The function $g(\hat z)$ also satisfies the
integral equation (\ref{equ:trimers:inteqgr}); this leads to another
similar set of conditions on the coefficients~$A_p$.

The form of the equations (\ref{equ:trimers:cancelnegq}) and
(\ref{equ:trimers:cancelposq}) and their analogue from
(\ref{equ:trimers:inteqgr}) suggests that for $\betaL$
and $\betaR$ fixed the coefficients $A_p$ should be
power series in~$\delta$,
\begin{equation}
  A_p = A_p^{(0)} + A_p^{(1)} \delta + A_p^{(2)} \delta^2 + \dots
  \label{equ:trimers:power}
\end{equation}
We would like to determine the coefficients~$A_p^{(h)}$.

When $t$ approaches the boundary of the annulus, $|t| \to \delta$ or
$|t| \to \delta^{-1}$, the unperturbed function $g(\hat z)$ given by
(\ref{equ:trimers:g2}) becomes of the order~$\delta$.  It seems
reasonable to assume that the terms $A_p t^p$ of the
perturbed function $g(\hat z)$ do not grow faster than this, so the
coefficients $A_p^{(h)}$ with $h < |p|-1$ must be zero.

Consider (\ref{equ:trimers:one}) and its analogue
from~(\ref{equ:trimers:inteqgr}).  Substitution of the power series
(\ref{equ:trimers:power}) yields, after rearrangement of the terms,
\begin{equation}
  \sum_{h=0}^{\infty} \delta^h
  \left( \sum_{p=-(h+1)}^{h+1} t_k^p A_p^{(h)} \right) = 1
  \qquad \mbox{for} \qquad k=1, 5.
  \label{equ:trimers:deltaparts}
\end{equation}
The $\delta^0$ part gives
\begin{displaymath}
  t_k A_1^{(0)} + t_k^{-1} A_{-1}^{(0)} = 1
  \qquad \mbox{for} \qquad k=1, 5.
\end{displaymath}
The unique solution of these equations reproduces the unperturbed
function $g(\hat z)$ given by~(\ref{equ:trimers:g2}).
For $1 \le h \le 3$ the $\delta^h$ part of
(\ref{equ:trimers:deltaparts}) gives
\begin{displaymath}
  t_k A_1^{(h)} + t_k^{-1} A_{-1}^{(h)} = 0
  \qquad \mbox{for} \qquad k=1, 5.
\end{displaymath}
This implies that
$A_1^{(h)}$ and $A_{-1}^{(h)}$ are zero.
The $\delta^4$ part of (\ref{equ:trimers:deltaparts}) gives
\begin{displaymath}
  t_k A_1^{(4)} + t_k^{-1} A_{-1}^{(4)} +
  t_k^5 A_5^{(4)} + t_k^{-5} A_{-5}^{(4)} = 0
  \qquad \mbox{for} \qquad k=1, 5.
\end{displaymath}
These equations have two linearly independent solutions, one of which
satisfies~(\ref{equ:trimers:posnegsymm}).  Substituting these results
into (\ref{equ:trimers:power}) and (\ref{equ:trimers:laurent}) yields
\begin{equation}
  g(\hat z) =
  t + t^{-1} + C \left(t^{-5} - t\right) +
  C^* \left(t^5 - t^{-1}\right) + O\!\left(\delta^5\right),
  \label{equ:trimers:gperturbed}
\end{equation}
where we have written
\begin{displaymath}
  A_{-5}^{(4)} \delta^4 = C
  \qquad \mbox{and} \qquad
  A_5^{(4)} \delta^4 = C^*.
\end{displaymath}
Note that (\ref{equ:trimers:gperturbed}) can be written in the
form~(\ref{equ:trimers:g}).
We have used the equations (\ref{equ:trimers:cancelnegq}) and
(\ref{equ:trimers:cancelposq}) only to come up with the series
expansion~(\ref{equ:trimers:power}).  These equations would be needed
if the coefficients $A_p^{(h)}$ with~$h>4$ were to be determined.
Knowledge of these coefficients would yield a solution to the integral
equations (\ref{equ:trimers:inteqgl}) and (\ref{equ:trimers:inteqgr})
also for a finite opening between $\hat \bL$
and~$\hat \bR$.
Unfortunately we have not been able to calculate these coefficients,
but fortunately they are not needed, because the present purpose is
only to compute
$\rhoL$, $\rhoR$, $\phiL$,
$\phiR$ and
$\Phi = -(\SigmaL + \SigmaR)$ only to leading
order in~$\delta$.

As our aim is to calculate the quantities $\rhoL$,
$\rhoR$, $\phiL$, $\phiR$
and $\Phi$, we substitute (\ref{equ:trimers:laurent}) and
(\ref{equ:trimers:power}) into the integral expressions
from Subsection~\ref{sub:trimers:physquant1}.  For
(\ref{equ:trimers:intrhol2}) this gives, after transforming to $t$ as
integration variable:
\begin{equation}
  \rhoL =
  \frac{1}{2 \pi \mathrm{i}} \sum_{p=-\infty}^{\infty}
  \sum_{h=|p|-1}^{\infty} A_p^{(h)} \delta^h
  \int_{\betaL \delta^{-1}}
      ^{\betaL \delta}
  t^p \frac{1}{\sqrt{\hat z^2 + 4}}
  \frac{\mathrm{d} \hat z}{\mathrm{d} t} \, \mathrm{d} t.
  \label{equ:trimers:intrhol3}
\end{equation}
For each $p$ and $h$ we determine the order in $\delta$ of the
contribution.  When $t \to 0$ or $t \to \infty$ the integrand is
proportional to $t^{p+5}$ and $t^{p-7}$, respectively.  Hence the
integral is bounded, of order 0 in $\delta$ that is, for $|p| \le 5$;
logarithmic in $\delta$ for $|p|=6$; of order $6-|p|$ in $\delta$
for~$|p| \ge 7$.  Note that the coefficients $A_p^{(h)}$ with $|p|=6$
are zero.  Let $m$ denote the order in $\delta$ of the integral.
The order in $\delta$ of the whole contribution is $h+m$:
\begin{displaymath}
  \begin{tabular}{||c||c|c|c||}
  \hline
  $|p|$   & $m$     & $h$         & $h+m$ \\
  \hline
  $1$     & $0$     & $0$         & $0$  \\
  $1$     & $0$     & $4$         & $4$  \\
  $1$     & $0$     & $\ge 5$     & $\ge 5$ \\
  \hline
  $5$     & $0$     & $4$         & $4$  \\
  $5$     & $0$     & $\ge 5$     & $\ge 5$ \\
  \hline
  $\ge 7$ & $6-|p|$ & $\ge |p|-1$ & $\ge 5$ \\
\hline
\end{tabular}
\end{displaymath}
Therefore $\rhoL$ in (\ref{equ:trimers:intrhol3}) has a
$\delta^0$ contribution from the unperturbed part in the RHS of
(\ref{equ:trimers:gperturbed}), a $\delta^4$ contribution from the
part involving $C$ and $C^*$, and contributions of higher order in
$\delta$ from the parts collected in the
$O\!\left(\delta^5\right)$ term, so
\begin{displaymath}
  \rhoL =
  \frac{1}{2 \pi \mathrm{i}}
  \int_{\betaL \delta^{-1}}
      ^{\betaL \delta}
  \left[t + t^{-1} + C \left(t^{-5} - t\right) +
        C^* \left(t^5 - t^{-1}\right)\right]
  \frac{1}{\sqrt{\hat z^2 + 4}}
  \frac{\mathrm{d} \hat z}{\mathrm{d} t} \, \mathrm{d} t +
  O\!\left(\delta^5\right).
\end{displaymath}
In the RHS the integration limits may be changed from
$\betaL \delta^{-1}$ and $\betaL \delta$
to $\infty$ and $0$, as this makes a difference
$O\!\left(\delta^5\right)$.
Transforming back to $z$ as integration variable then gives
\begin{displaymath}
  \rhoL =
  \frac{1}{2 \pi \mathrm{i}} \int_{\Xi^{(0)}}
  \left[t + t^{-1} + C \left(t^{-5} - t\right) +
        C^* \left(t^5 - t^{-1}\right)\right] \,
  \mathrm{d} z + O\!\left(\delta^5\right),
\end{displaymath}
where $\Xi^{(0)}$ denotes the unperturbed ($\delta = 0$)
contour.  Hence $\rhoL$ is given to leading order in
$C \sim \delta^4$ by (\ref{equ:trimers:intrhol2}), where now
$f(z)$ is given by (\ref{equ:trimers:fperturbed}) instead of
(\ref{equ:trimers:f}), and integration is over the unperturbed contour.
Note that $\betaL$ and $\betaR$
do not occur in this expression.
Similar arguments show that fully analogous results hold for
$\rhoR$, $\phiL$, $\phiR$,
$\SigmaL$ and~$\SigmaR$: up to
$O\!\left(\delta^5\right)$ they are given by
the integrals (\ref{equ:trimers:intrhol2}), (\ref{equ:trimers:intphil})
and (\ref{equ:trimers:intSIGL}) or their R-analogues, with $f(z)$ given
by~(\ref{equ:trimers:fperturbed}).  Therefore we have now obtained
these quantities to leading order, namely $\delta^4$, in the
parameter $\delta$ that describes the infinitesimally small opening
$\hat \bL - \hat \bR \sim \delta^6$ between the
end points of $\hat \Xi$ and~$\hat \Eta$.

\end{document}